\documentclass[twocolumn,showpacs,amsmath,amssymb,superscriptaddress]{revtex4-1}
\usepackage{dcolumn}
\usepackage{color}
\usepackage{graphicx}
\usepackage{amsmath}
\usepackage{multirow}
\usepackage{bm}
\usepackage{url}

\begin{document}

\title{Shape transitions in odd-mass $\gamma$-soft nuclei within
the interacting boson-fermion model based on the Gogny energy density functional}

\author{K.~Nomura}
\affiliation{Physics Department, Faculty of Science, University of
Zagreb, HR-10000 Zagreb, Croatia}
\affiliation{Center for Computational
Sciences, University of Tsukuba, Tsukuba 305-8577, Japan}

\author{R.~Rodr\'iguez-Guzm\'an}
\affiliation{Physics Department, Kuwait University, 13060 Kuwait, Kuwait}

\author{L.~M.~Robledo}
\affiliation{Departamento de F\'\i sica Te\'orica, Universidad
Aut\'onoma de Madrid, E-28049 Madrid, Spain}

\affiliation{Center for Computational Simulation,
Universidad Polit\'ecnica de Madrid,
Campus de Montegancedo, Boadilla del Monte, 28660-Madrid.
}

\date{\today}

\begin{abstract}
The interacting boson-fermion model (IBFM), with parameters determined
 from the microscopic Hartree-Fock-Bogoliubov (HFB) approximation, based 
 on the parametrization D1M of the Gogny energy density 
 functional (EDF), is employed to
 study the structural evolution in odd-mass $\gamma$-soft nuclei.
 The deformation energy surfaces of even-even nuclei, single-particle energies
 and occupation  probabilities of the corresponding odd-mass systems have been
 obtained within the constrained HFB approach. Those building blocks are then 
 used  as a microscopic input to build the IBFM Hamiltonian. 
 The coupling constants of the boson-fermion interaction terms are
 taken as free parameters, fitted to reproduce experimental low-lying spectra.
 The diagonalization 
 of the IBFM Hamiltonian provides the spectroscopic properties 
 for the studied odd-mass nuclei. The procedure has been applied to compute 
 low-energy excitation spectra and
 electromagnetic transition rates, in the case of the
 $\gamma$-soft odd-mass systems
 $^{129-137}$Ba, $^{127-135}$Xe,  $^{129-137}$La and $^{127-135}$Cs. 
 The  calculations provide a reasonable agreement with the available experimental data
 and agree well with previous results based on the relativistic mean-field 
 approximation. 
\end{abstract}

\keywords{}

\maketitle

% ----------------------------------------------------------------------

\section{Introduction}

% ----------------------------------------------------------------------

The microscopic description of odd-mass nuclei is one of the most 
challenging topics in today's low-energy nuclear structure physics 
\cite{BM}. In those systems, the interplay between the single-particle 
and collective degrees of freedom plays a key role. Among the 
theoretical approaches that are nowadays becoming a standard tool for 
microscopic nuclear structure studies are, the large-scale Shell Model 
(SM) \cite{caurier2005} and self-consistent mean-field (SCMF) methods 
based on  nuclear energy density functionals (EDFs) 
\cite{RS,bender2003,vretenar2005,niksic2011}. These theoretical 
approaches have already been successfully employed in the study of the  
properties of even-even nuclei. However, a systematic investigation of 
odd-mass nuclei, particularly in the medium- and heavy-mass regions of 
the nuclear chart,  still represents a major computational challenge.

On the one hand, the SM framework includes the relevant nuclear 
correlations  via the diagonalization of the effective Hamiltonian, 
defined in a given valence space, and  allows a direct access to 
spectroscopic properties. However, in heavy and/or open-shell nuclei, 
the dimension of the SM Hamiltonian matrix becomes too large so as to 
make the approach computationally  less feasible. On the other hand, 
the SCMF methods  allow  a global description of the intrinsic 
properties all over the nuclear chart. Nevertheless, in order to 
describe spectroscopic properties, one has to go beyond the mean-field 
level to include dynamical correlations arising from the restoration of 
the (spontaneously)  broken symmetries  and/or  fluctuations in the 
collective parameters. This can be done  within the framework the 
symmetry-projected generator coordinate method (GCM) 
\cite{RS,rayner2002,bender2003,niksic2011}. For odd-mass nuclei, 
however, the GCM framework should be extended to explicitly take into 
account both the time-odd fields and the effect of blocking at each 
deformation on the energy surface of a given nuclear system 
\cite{bally2014}. Such a calculation, however, tends to be highly 
demanding from a computational point of view and indeed has been mainly 
limited to light nuclei \cite{bally2014,borrajo2016}.

The computational difficulties already mentioned, suggest the 
exploration of alternative  schemes that can be used for odd-mass 
nuclei with arbitrary masses. Examples of such approaches are, the 
particle-vibration coupling scheme \cite{BM,bohr1953} based on both 
nonrelativistic \cite{mizuyama2012,tarpanov2014a,tarpanov2014b,niu2015} 
and relativistic \cite{litvinova2007,litvinova2011} EDFs as well as  
the symmetry-based approaches 
\cite{iachello1979,scholten1985,IBFM,iachello2011,petrellis2011}. 
Here, we also mention other fully-microscopic approaches
to odd-nucleon systems such as coupled-cluster calculations \cite{jansen2016}
and the equation of motion phonon method
\cite{degregorio2016,degregorio2017}. Such approaches resort 
to realistic nucleon-nucleon
interactions though they have so far been limited to light nuclear systems.

Recently, a method has  been developed  to describe odd-mass nuclear 
systems \cite{nomura2016odd}, based  on the nuclear EDF framework 
and the particle-core coupling scheme. In this case, the even-even core 
is described in terms of bosonic degrees of freedom, i.e., the 
interacting boson model (IBM) \cite{IBM}, and the particle-boson 
coupling is taken into account within the interacting boson-fermion 
model (IBFM) \cite{IBFM}. The quadrupole deformation energy surfaces 
for the even-even boson-core nucleus, the single-particle energies as 
well as the  occupation numbers for the considered odd-mass system are 
all computed  within the SCMF method based on a given nuclear EDF. They  
are then used as a microscopic input to determine the boson-core 
Hamiltonian and the boson-fermion interaction terms. The coupling 
constants of the boson-fermion interaction are the only free 
parameters, which are  adjusted to selected spectroscopic data. The 
method has already been tested for axially-deformed odd-mass Sm and Eu  
\cite{nomura2016odd,nomura2016qpt,nomura2017odd-2} and  $\gamma$-soft 
odd-mass Ba, Xe, La and Cs nuclei in the $A\approx 130$ mass region 
\cite{nomura2017odd-1} using the relativistic DD-PC1 \cite{DDPC1} and 
the Gogny EDFs. It has been shown, that the method works equally well 
regardless of the underlying EDF used as microscopic input.

In this paper, we consider the structural evolution between 
nearly spherical and $\gamma$-soft shapes in the case of 
odd-mass  Ba, Xe, La and Cs  nuclei  with $A\approx 130$. 
As will be shown later on in this paper  our results,  based  on the Gogny-EDF, exhibit 
a similar level of agreement with the experiment as the ones obtained 
in Ref.~\cite{nomura2017odd-1} corroborating the validity of the method
also for $\gamma$-soft nuclei. Once again, both relativistic and/or non-relativistic
EDF produce equivalent results.

Until now, a number of purely phenomenological IBFM calculations have 
already been performed in this mass region 
\cite{alonso1984,arias1985,cunningham1982a,cunningham1982b,alonso1987,dellagiacoma1988phdthesis,ABUMUSLEH2014}. 
A virtue of the present approach is that, though its applicability is 
currently limited to nuclei where spectroscopic data are available, the 
boson-core Hamiltonian, single-particle energies and occupation 
probabilities of the considered odd-mass systems are completely 
determined by  fully microscopic SCMF calculations. We compare our 
results with those obtained in the phenomenological IBFM studies 
already mentioned.

The paper is organized as follows. In Sec.~\ref{sec:model}, we briefly 
outline the method employed and discuss the  parameters of the IBFM 
Hamiltonian. The results for the even-even boson core nuclei 
$^{128-136}$Ba and $^{126-134}$Xe are presented in 
Sec.~\ref{sec:even-results}. In particular, we discuss the 
corresponding mean-field and (mapped) IBM energy surfaces as well as 
the low-energy excitation spectra. The spectroscopic properties of the 
odd-mass nuclei $^{129-137}$Ba, $^{127-135}$Xe, $^{129-137}$La and 
$^{127-135}$Cs are presented in Sec.~\ref{sec:odd-results} where, the 
computed low-energy positive- and negative-parity excitation spectra, 
$B(E2)$ transition strengths, spectroscopic quadrupole and magnetic 
moments are discussed and compared with  the available experimental 
data. We end up this section with a detailed analysis of the excitation 
spectrum and electromagnetic properties for a few selected nuclei for 
which experimental data are available. Finally, Sec.~\ref{sec:summary} 
is devoted to the conclusions and work perspectives.

% ----------------------------------------------------------------------

\section{Description of the model\label{sec:model}}

% ----------------------------------------------------------------------

In this section, we briefly outline the theoretical  framework 
used in this study \cite{nomura2016odd}. We also 
discuss the parameters of the IBFM Hamiltonian employed in the 
calculations.

The IBFM Hamiltonian, used to describe the studied odd-mass  nuclei, 
consists of three terms, i.e, the even-even boson core or 
Interacting Boson Model (IBM) Hamiltonian $\hat H_B$, the 
single-particle Hamiltonian for unpaired fermions $\hat H_F$ and the 
boson-fermion coupling term $\hat H_{BF}$
\begin{equation}
\label{eq:ham}
 \hat H=\hat H_B + \hat H_F + \hat H_{BF}. 
\end{equation}

The building blocks of the IBM are the $s$ and $d$ bosons, which
represent the collective pairs of valence nucleons \cite{OAI} coupled
to angular momentum $J^\pi=0^+$ and $2^+$, respectively. 
The number of bosons $N_B$ and fermions $N_F$ are assumed to be
conserved separately. 
We restrict ourselves to the simplest case
$N_F=1$, where contributions from three or higher
quasiparticle configurations are not included. 
In addition, no distinction is made between neutron and 
proton bosons. 

The IBM Hamiltonian $\hat H_B$ reads
\begin{equation}
\label{eq:ibm}
 \hat H_B = \epsilon_d\hat n_d + \kappa\hat Q_B\cdot\hat Q_B, 
\end{equation}
and is given in terms of the $d$-boson number operator $\hat n_d=d^{\dagger}\cdot\tilde d$,
and the quadrupole operator $\hat Q_B=s^{\dagger}\tilde d+d^{\dagger}\tilde s +
\chi[d^{\dagger}\times\tilde d]^{(2)}$. The 
quantities $\epsilon_d$, $\kappa$, and $\chi$ 
represent parameters of the Hamiltonian $\hat H_{B}$.

On the other hand, the single-fermion Hamiltonian takes the form
\begin{equation}
\hat
H_F=\sum_{j}\epsilon_j[a^{\dagger}_j\times\tilde a_j]^{(0)} 
\end{equation} 
where $a^{\dagger}_j$ is the fermion creation
operator for the orbital $j$ and $\epsilon_j$ stands 
for the corresponding single-particle energy.

For the boson-fermion coupling Hamiltonian $\hat H_{BF}$ we have employed the  
simplest possible form that has been shown to be most relevant in the
phenomenological studies of Ref \cite{IBFM}: 
\begin{eqnarray}
\label{eq:bf}
 \hat H_{BF}
=&&\sum_{jj^{\prime}}\Gamma_{jj^{\prime}}\hat
  Q_B\cdot[a^{\dagger}_j\times\tilde a_{j^{\prime}}]^{(2)} \nonumber \\
&&+\sum_{jj^{\prime}j^{\prime\prime}}\Lambda_{jj^{\prime}}^{j^{\prime\prime}}
:[[d^{\dagger}\times\tilde a_{j}]^{(j^{\prime\prime})}
\times
[a^{\dagger}_{j^{\prime}}\times\tilde d]^{(j^{\prime\prime})}]^{(0)}:
\nonumber \\
&&+\sum_j A_j[a^{\dagger}\times\tilde a_{j}]^{(0)}\hat n_d.
\end{eqnarray}
The first, second and third terms are referred to as the dynamical
quadrupole, exchange, and monopole interactions, respectively. 
For the strength parameters $\Gamma_{jj^{\prime}}$,
$\Lambda_{jj^{\prime}}^{j^{\prime\prime}}$ and $A_j$ we have used 
the following expressions which were  derived within the generalized
seniority scheme \cite{scholten1985}: 
\begin{eqnarray}
\label{eq:dynamical}
&&\Gamma_{jj^{\prime}}=\Gamma_0\gamma_{jj^{\prime}} \\
\label{eq:exchange}
&&\Lambda_{jj^{\prime}}^{j^{\prime\prime}}=-2\Lambda_0\sqrt{\frac{5}{2j^{\prime\prime}+1}}\beta
_{jj^{\prime\prime}}\beta_{j^{\prime}j^{\prime\prime}} \\
\label{eq:monopole}
&&A_j=-A_0\sqrt{2j+1}
\end{eqnarray}
The quantities
$\gamma_{jj^{\prime}}=(u_ju_{j^{\prime}}-v_jv_{j^{\prime}})Q_{jj^{\prime}}$
and 
$\beta_{jj^{\prime}}=(u_jv_{j^{\prime}}+v_ju_{j^{\prime}})Q_{jj^{\prime}}$,
are given in terms of the occupation probabilities $u_j$ and $v_j$ 
for the orbital $j$ (satisfying  $u_j^2+v_j^2=1$) and
the matrix element of the quadrupole operator in the single-particle
basis $Q_{jj^{\prime}}=\langle j||Y^{(2)}||j^{\prime}\rangle$.
Furthermore, $\Gamma_0$, $\Lambda_0$ and $A_0$ denote 
strength parameters.
For a detailed account of the 
formulas in Eqs.~(\ref{eq:bf})-(\ref{eq:monopole}), as
well as a discussion of relevant applications to  odd-mass nuclei, the
reader is referred to Ref.~\cite{scholten1985}.

The first step to build the full IBFM Hamiltonian, $\hat H$  
Eq.~(\ref{eq:ham}), is to fix the parameters of the IBM Hamiltonian 
$\hat H_B$ by using the fermion-to-boson mapping procedure developed in 
Refs.~\cite{nomura2008,nomura2010}. Within this context, the fermionic 
($\beta$,$\gamma$)-deformation energy surface, obtained via mean-field 
calculations based on the parametrization D1M \cite{D1M} of the 
Gogny-EDF, is mapped onto the expectation value of $\hat H_B$ in the 
boson condensate state as defined in Ref \cite{ginocchio1980}. This 
procedure completely determines the parameters $\epsilon_d$, $\kappa$ 
and $\chi$.  Calculations have also been carried out with the 
parametrization D1S \cite{D1S} of the Gogny-EDF. However, as the 
results turn out to be rather similar to the ones provided by the 
Gogny-D1M EDF, we will just discuss the latter in this work.

For a more detailed account of the constrained Gogny-HFB approximation 
the reader is referred to Refs.~\cite{robledo2008,rayner2010pt}. 
Details of the fermion-to-boson mapping procedure for even-even system 
can be found in Refs.~\cite{nomura2008,nomura2010}. The parameters 
derived for the considered boson-core nuclei $^{128-136}$Ba and 
$^{126-134}$Xe are listed in Table~\ref{tab:paraB}.

% ----------------------------------------------------------------------
%                                                      T a b l e   I
% ----------------------------------------------------------------------
\begin{table}[hb!]
\caption{\label{tab:paraB} Parameters of the boson Hamiltonian $\hat
 H_B$ for  $^{128-136}$Ba and
 $^{126-134}$Xe. 
 The values of $\epsilon_d$ and $\kappa$ are in  MeV, while 
 $\chi$ is dimensionless. }
\begin{center}
\begin{tabular*}{\columnwidth}{p{2.0cm}p{2.0cm}p{2.0cm}p{2.0cm}}
\hline\hline
\textrm{} &
\textrm{$\epsilon_{d}$}&
\textrm{$\kappa$}&
\textrm{$\chi$} \\
\hline
$^{128}$Ba & 0.120 & -0.080 & -0.14 \\
$^{130}$Ba & 0.150 & -0.081 & -0.16 \\
$^{132}$Ba & 0.265 & -0.081 & -0.10 \\
$^{134}$Ba & 0.620 & -0.083 & -0.37 \\
$^{136}$Ba & 1.000 & -0.090 & -0.80 \\
$^{126}$Xe & 0.245 & -0.079 & -0.13 \\
$^{128}$Xe & 0.280 & -0.079 & -0.22 \\
$^{130}$Xe & 0.375 & -0.081 & -0.12 \\
$^{132}$Xe & 0.620 & -0.086 & -0.30 \\
$^{134}$Xe & 1.000 & -0.090 & -0.87 \\
\hline\hline
\end{tabular*}
\end{center}
\end{table}

\begin{figure}[htb!]
\begin{center}
\includegraphics[width=\linewidth]{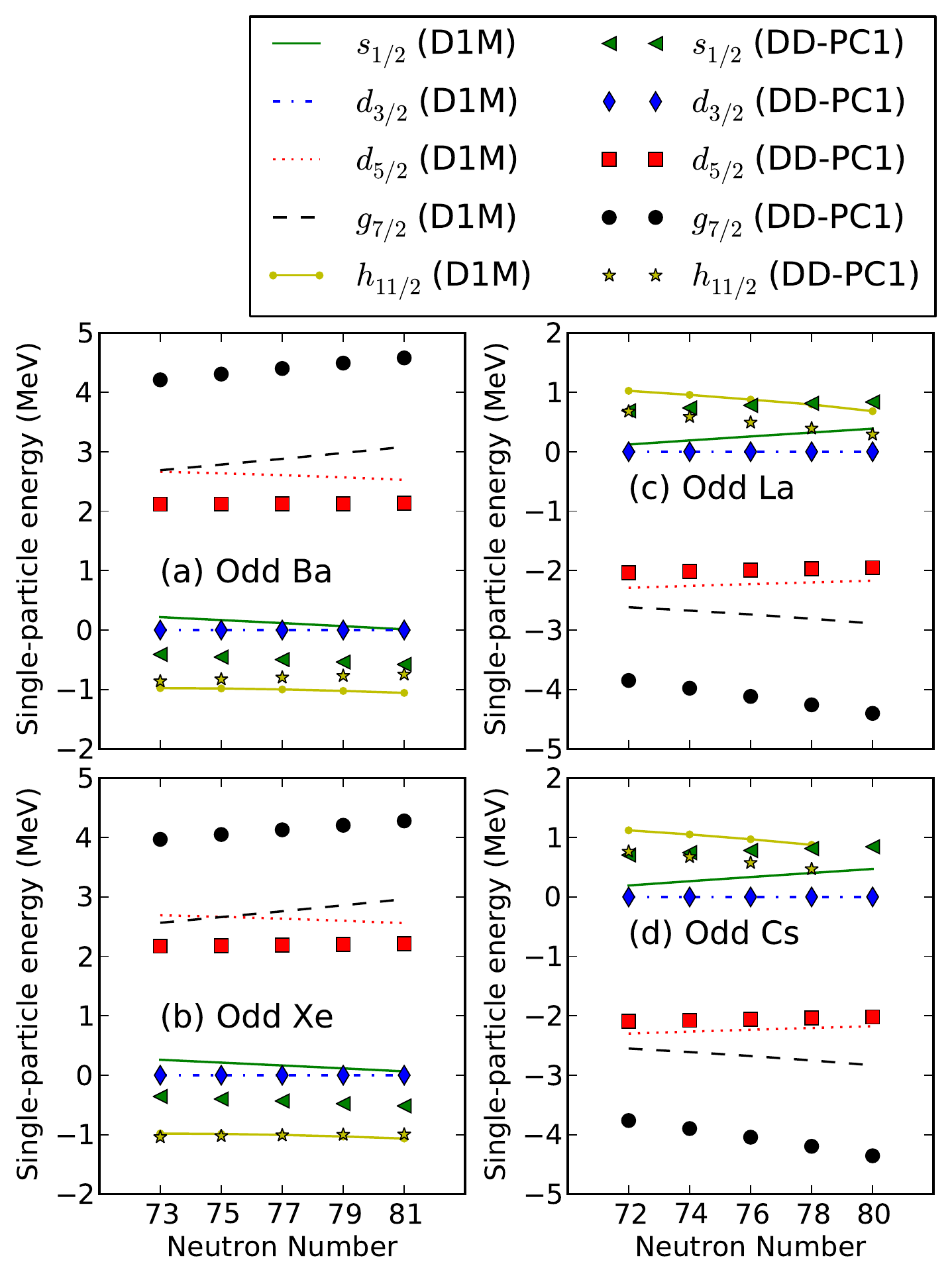}
\caption{(Color online) The single-particle energies corresponding to the $3s_{1/2}$,
 $2d_{3/2}$, $2d_{5/2}$, $1g_{7/2}$ and $1h_{11/2}$ orbitals employed in the present
 study (denoted as ``D1M'') for the considered odd-mass nuclei
 are plotted with respect to the $2d_{3/2}$ single-particle level. 
 Results from Ref.~\cite{nomura2017odd-1} are also included in the plot (denoted as ``DD-PC1'').
 } 
\label{fig:spe}
\end{center}
\end{figure}

\begin{figure}[htb!]
\begin{center}
\includegraphics[width=\linewidth]{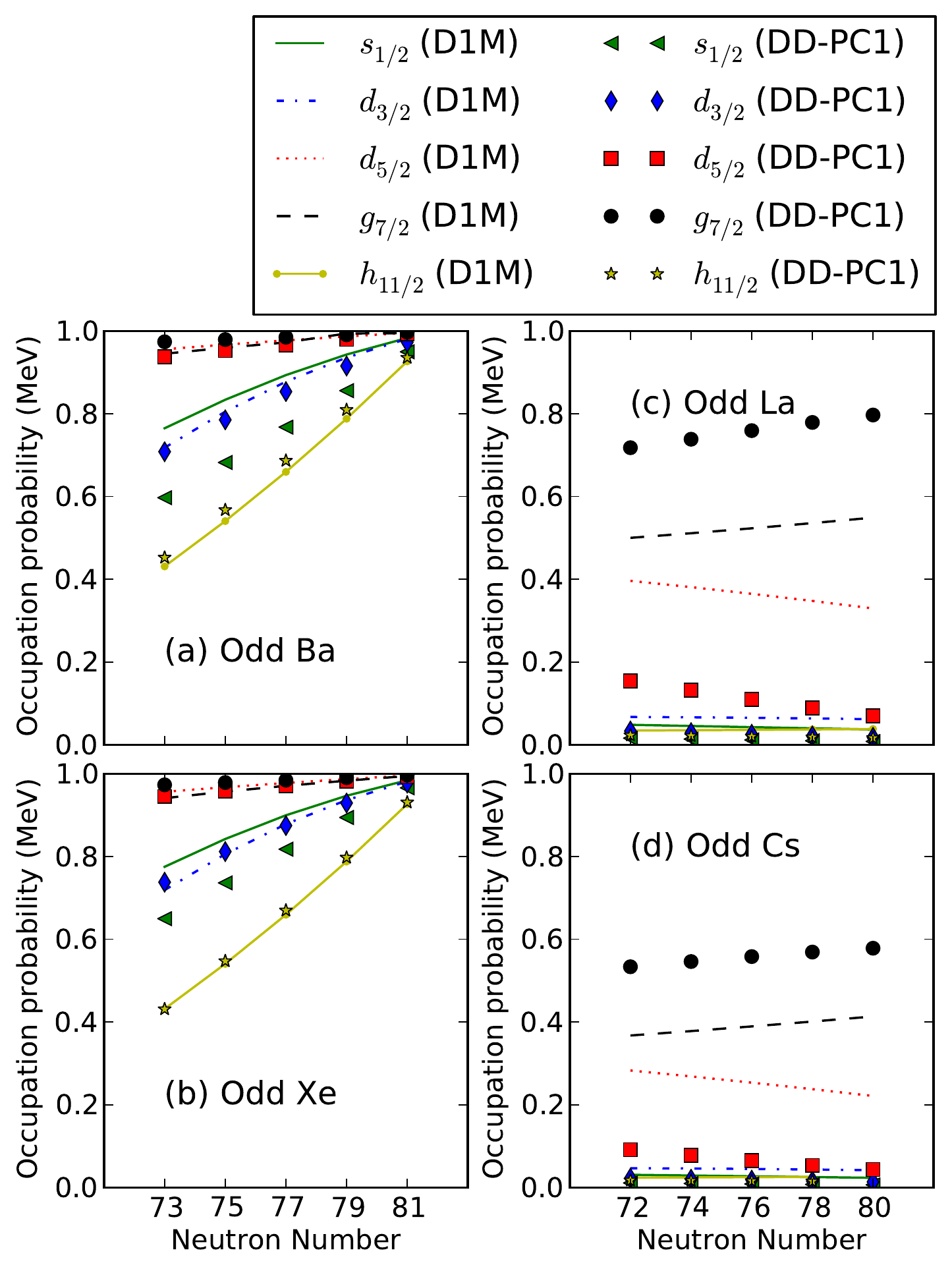}
\caption{(Color online) The same as in  Fig.~\ref{fig:spe}, but for
 occupation probabilities of the $3s_{1/2}$, 
 $2d_{3/2}$, $2d_{5/2}$, $1g_{7/2}$ and $1h_{11/2}$ orbitals.} 
\label{fig:vv}
\end{center}
\end{figure}

For all the nuclei considered in this work 
$^{129-137}$Ba, $^{127-135}$Xe, $^{129-137}$La and $^{127-135}$Cs
we have considered as the fermion valence space (neutrons for Ba and Xe and
protons for La and Cs), all the spherical 
single-particle orbitals between magic numbers 50 and 82, i.e., 
$3s_{1/2}$, $2d_{3/2}$, $2d_{5/2}$ and $1g_{7/2}$ for positive-parity 
states and $1h_{11/2}$ for negative-parity states. The 
single-particle energies $\epsilon_j$ and the occupation probabilities 
$v_j^2$ are obtained from   Gogny-D1M HFB calculations 
at the spherical configuration \cite{nomura2016odd}. In those calculations, for a given 
odd-mass nucleus with the odd neutron (proton) number $N_{0}$ 
($Z_{0}$), the standard even number parity constrained Gogny-HFB 
approach (i.e., without blocking) has been employed but using $N_{0}$ 
($Z_{0}$) for the neutron (proton) number constraint. The 
single-particle energies and occupation probabilities obtained for the 
considered odd-$A$ nuclei, are shown in Figs.~\ref{fig:spe} and 
\ref{fig:vv}, respectively.

% ----------------------------------------------------------------------
%                                            F i g u r e   1
% ----------------------------------------------------------------------

% ----------------------------------------------------------------------
%                                            F i g u r e   2
% ----------------------------------------------------------------------

% ----------------------------------------------------------------------
%                                            F i g u r e   3
% ----------------------------------------------------------------------

\begin{figure}[htb!]
\begin{center}
\includegraphics[width=\linewidth]{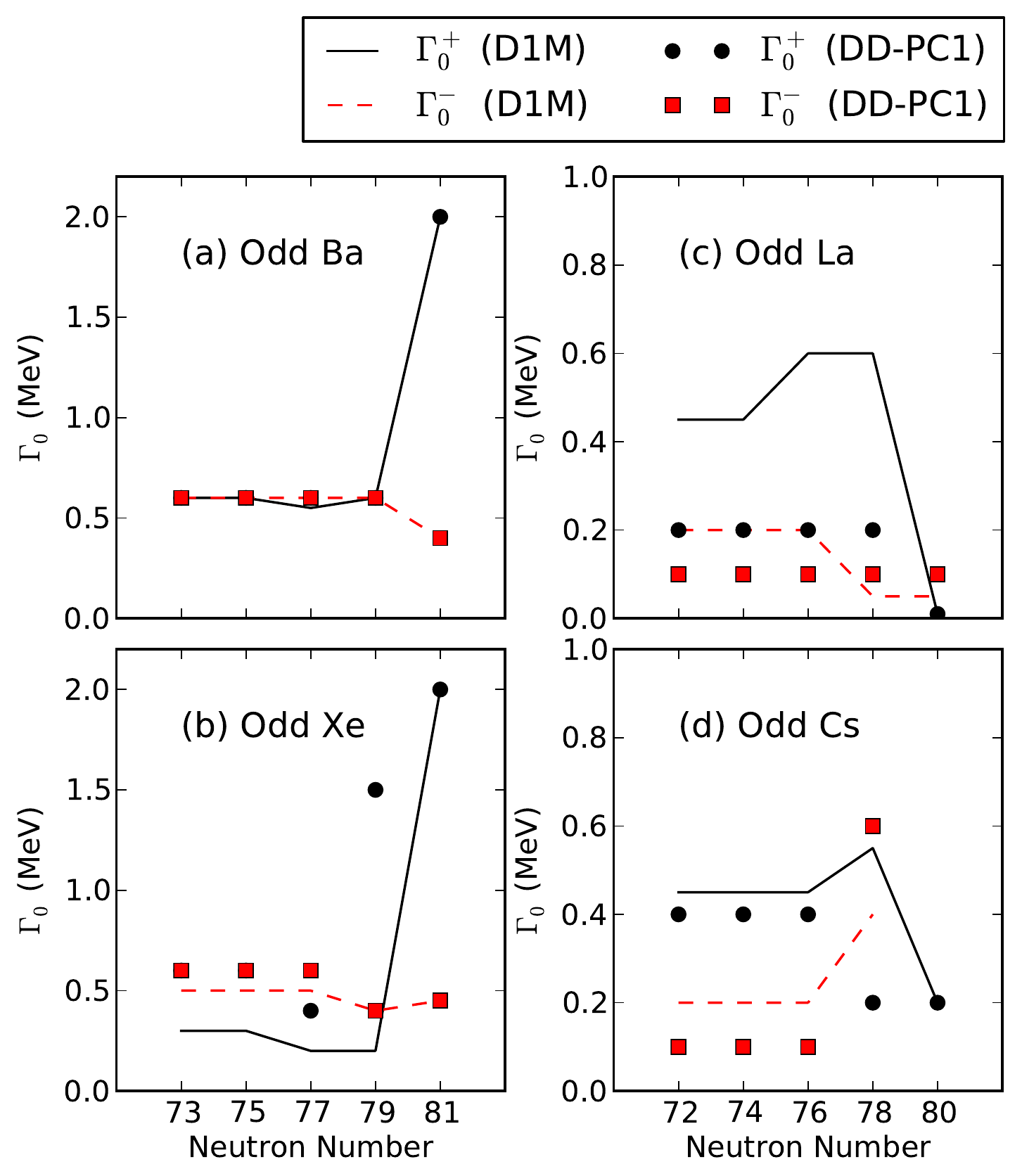}
\caption{(Color online) The strength parameter $\Gamma_0$ 
 Eq.~(\ref{eq:dynamical})  (denoted by ``D1M'') is plotted
  for both positive- ($\Gamma_0^+$) and
 negative-parity ($\Gamma_0^-$) states.
 Results from Ref.~\cite{nomura2017odd-1} are also included in the plot (denoted as ``DD-PC1'').
 } 
\label{fig:para-dyn}
\end{center}
\end{figure}

% ----------------------------------------------------------------------
%                                            F i g u r e   4
% ----------------------------------------------------------------------

\begin{figure}[htb!]
\begin{center}
\includegraphics[width=\linewidth]{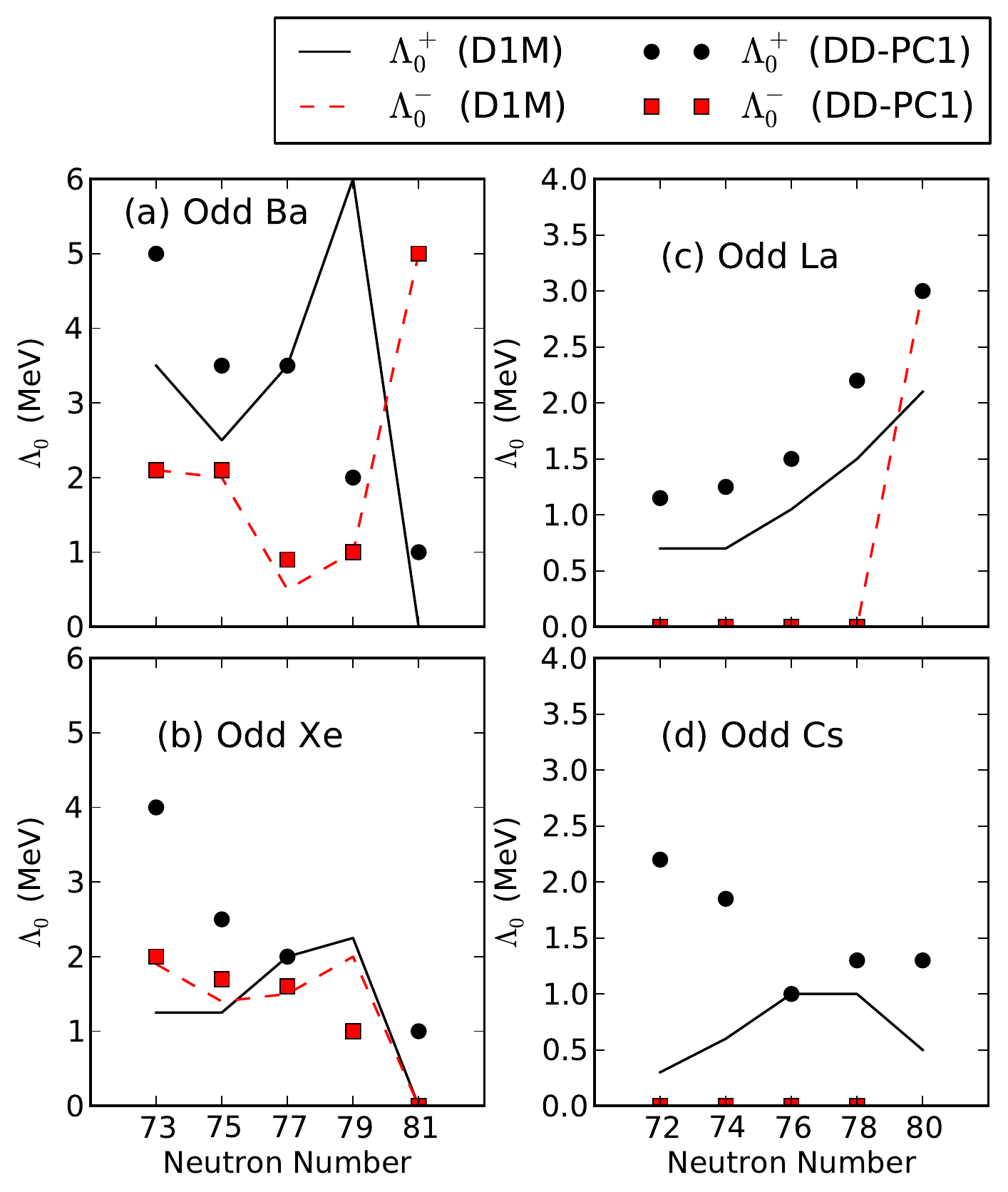}
\caption{(Color online) The same as in Fig.~\ref{fig:para-dyn}, but for the
 strength parameter $\Lambda_0$ of the exchange term  Eq.~(\ref{eq:exchange}).} 
\label{fig:para-exc}
\end{center}
\end{figure}

% ----------------------------------------------------------------------
%                                                      T a b l e   I I
% ----------------------------------------------------------------------

\begin{table}[hb!]
\caption{\label{tab:monopole} The strength parameters of the monopole
 terms $A_j^{\prime}$. All entries
 are in  MeV. }
\begin{center}
\begin{tabular*}{\columnwidth}{p{1.33cm}p{1.33cm}p{1.33cm}p{1.33cm}p{1.33cm}p{1.33cm}}
\hline\hline
\textrm{} &
\textrm{$A_{1/2}^{\prime}$} &
\textrm{$A_{3/2}^{\prime}$}&
\textrm{$A_{5/2}^{\prime}$}&
\textrm{$A_{7/2}^{\prime}$} &
\textrm{$A_{11/2}^{\prime}$} \\
\hline
$^{129}$Ba & -0.23 &  &  & -0.58 &  \\
$^{131}$Ba &  &  &  &  &  \\
$^{133}$Ba &  &  &  &  &  \\
$^{135}$Ba &  & -0.35 &  &  & -0.5 \\
$^{137}$Ba &  & -1.2 &  &  & -0.3 \\
$^{127}$Xe & -0.22 &  &  & -0.64 &  \\
$^{129}$Xe & -0.10 &  &  &  &  \\
$^{131}$Xe &  &  &  &  &  \\
$^{133}$Xe &  & -0.25 &  &  & -0.15 \\
$^{135}$Xe &  & -1.5 &  &  & -0.15 \\
$^{129}$La &  &  & -0.35 &  &-0.40 \\
$^{131}$La &  &  & -0.40 &  &  \\
$^{133}$La &  &  & -0.22 &  &  \\
$^{135}$La &  &  & -0.25 &  &  \\
$^{137}$La &  &  &  &  & -0.10 \\
$^{127}$Cs &  &  & -0.05 &  &  \\
$^{129}$Cs &  &  & -0.94 &  &   \\
$^{131}$Cs &  &  & -0.25 &  &   \\
$^{133}$Cs &  &  & -0.30 &  & -0.35 \\
$^{135}$Cs &  &  & -0.90 &  & \\
\hline\hline
\end{tabular*}
\end{center}
\end{table}

The coupling constants of the
boson-fermion interaction term $\hat H_{BF}$ ($\Gamma_0$, $\Lambda_0$
and $A_0$) are treated as free parameters. They are fitted, for each 
nucleus and separately for positive- and negative-parity states, to
a few experimental excitation spectra 
\cite{nomura2016odd}. The single-particle energies calculated within
the Gogny-HFB method are somewhat at variance with those employed in
purely phenomenological IBFM studies in this mass region  
\cite{cunningham1982a,cunningham1982b,dellagiacoma1988phdthesis}. 
For instance, in Ref.~\cite{dellagiacoma1988phdthesis} for the odd-$A$ Ba and Xe nuclei, the
$3s_{1/2}$ orbital lies around 0.4 MeV below the
$2d_{3/2}$ orbital. In the present study, on the other hand, the 
two orbitals are nearly degenerated as can be seen from
Figs.~\ref{fig:spe}(a) and \ref{fig:spe}(b). 
In order to better reproduce the ordering of the energy levels, we have
assumed that the monopole strength parameter for the positive-parity states, denoted by $A^+_0$, is
allowed to be $j$ dependent, i.e., $A_0^+\equiv A_j^{\prime}$.  
A similar  assumption was made in the previous study \cite{nomura2017odd-1}.

The fitted strength parameters of the boson-fermion interaction terms
$\Gamma_0^{\pm}$ and $\Lambda_0^{\pm}$ are plotted in Figs.~\ref{fig:para-dyn} and
\ref{fig:para-exc}, respectively. On the other hand, the monopole
strength parameters $A^{\prime}_j$ are listed in
Table~\ref{tab:monopole}. 
The behavior of some of the fitted parameters, as functions of the
nucleon number, reflects structural changes along a given isotopic
chain. For instance, one observes kinks in the $\Gamma_0^+$ values for the odd-$A$ La and Cs 
around the neutron number $N=78$ (Figs.~\ref{fig:para-dyn}(c) and \ref{fig:para-dyn}(d)) and
in  the
$\Lambda_0^+$ ones for the odd-$A$ Ba and Xe around $N=79$ (Fig.~\ref{fig:para-exc}(a)
and \ref{fig:para-exc}(b)). 
As will be shown, this is supported by the deformation energy surfaces for the
 boson-core nuclei
with those neutron numbers exhibiting flat-bottomed potential
characteristic of transitional systems (see, Figs.~\ref{fig:pes-d1m} and \ref{fig:pes-mapped}). 

Here, we briefly summarize some  differences and similarities between 
the IBFM parameters employed in the present study and those obtained in 
Ref.~\cite{nomura2017odd-1}:
 \begin{itemize}

\item Major differences are found in the parameters $\epsilon_d$
      and $\kappa$ of the IBM Hamiltonian obtained in this work
      and the ones of Ref.~\cite{nomura2017odd-1}. In particular, 
      the value of $\epsilon_d$ ($\kappa$) used in
      Ref.~\cite{nomura2017odd-1} is significantly smaller (larger)
      than the one obtained in this work. To a large extent, these differences 
      reflect the  ones between the Gogny-D1M and DD-PC1 energy surfaces. 

\item From the comparison of the single-particle energies in Figs.~\ref{fig:spe}(a) 
      and \ref{fig:spe}(b), one realizes that the 
      ordering of the  $3s_{1/2}$ and $2d_{3/2}$ orbitals is different for the 
      Gogny-D1M and the relativistic DD-PC1 EDFs.  In addition, all five 
      Gogny-D1M single-particle levels lie closer to each other than 
      those provided by the DD-PC1 EDF \cite{nomura2017odd-1}. 

\item As seen from Fig.~\ref{fig:vv}(a) and \ref{fig:vv}(b), the occupation
      probabilities $v^2_j$ for the odd-$A$ Ba and Xe nuclei are similar for both
      Gogny-D1M and DD-PC1 results, except for the $v^2_{s_{1/2}}$
      values. On the other hand, for the odd-$A$ La (Fig.~\ref{fig:vv}(c)) and 
      Cs (Fig.~\ref{fig:vv}(d))
      nuclei, the values of $v^2_{g_{7/2}}$ and
      $v^2_{d_{5/2}}$ are substantially
      different from those obtained with  the DD-PC1 EDF.

\item As can be seen from Figs.~\ref{fig:para-dyn} and \ref{fig:para-exc}, 
      similar values are obtained for the  parameters $\Gamma_0^-$ and 
      $\Lambda_0^-$  as compared to the ones in Ref.~\cite{nomura2017odd-1}.      
      There are, however, considerable differences in the  $\Gamma_0^+$ 
      and $\Lambda_0^+$ values. This reflects  the quantitative differences 
      in the $\epsilon_j$ and $v^2_j$ values provided by the  Gogny-D1M 
      and DD-PC1 approaches. On the other hand, the $A_j^{\prime}$ values 
      are generally smaller in magnitude than those in Ref.~\cite{nomura2017odd-1} 
      (see Table~\ref{tab:monopole} and Tables IV and V in Ref.~\cite{nomura2017odd-1}). 

\end{itemize}

Once all the  building blocks of the IBFM Hamiltonian $\hat H$  
Eq.~(\ref{eq:ham}) are determined, this Hamiltonian is diagonalized in 
the spherical basis $|j, L, \alpha, J\rangle$ \cite{PBOS}, where $L$ 
and $J$ stand for the angular momenta of the boson and boson-fermion 
systems, respectively. On the other hand, $\alpha$ represents a set of 
U(5) quantum numbers \cite{IBM}. Note, that the   selection rule 
$|L-j|\leq J\leq L+j$ must be satisfied. The wave functions resulting 
from the diagonalization of $\hat H$ are  then used to compute 
electromagnetic properties, such as electric quadrupole (E2) and 
magnetic dipole (M1) transitions. 

The E2 transition operator reads 
\begin{equation}
\label{eq:TE2}
\hat T^{(E2)}=\hat e_B\hat Q_B-e_F\sum_{jj^{\prime}}
\frac{1}{\sqrt{5}}
\gamma_{jj^{\prime}}
[a^{\dagger}\times\tilde a_{j^{\prime}}]^{(2)},
\end{equation}
where the first and second terms represent the boson and fermion E2 
operators, respectively. In Eq.~(\ref{eq:TE2}), $\hat Q_B$ is the 
quadrupole operator already defined in Eq.~(\ref{eq:ibm}) with the same 
value of the parameter $\chi$. The effective bosonic charge $e_B$ is 
fitted, for each nucleus, to reproduce the experimental $B(E2; 
2^+_1\rightarrow 0^+_1)$ value of the corresponding even-even 
boson-core. On the other hand, the effective fermionic charge  $e_F$ is 
taken separately for proton ($\pi$) and neutron ($\nu$) as 
$e_F^{\pi}=0.25\,e$b and $e_F^{\nu}=0.125\,e$b \cite{nomura2017odd-1}.

The M1 transition operator is given by
\begin{equation}
 \hat T^{(M1)}=\sqrt{\frac{3}{4\pi}}(\hat T^{(M1)}_B + \hat T^{(M1)}_F).
\end{equation}
where $\hat T^{(M1)}_B=g_B\hat L$, with $\hat 
L=\sqrt{10}[d^{\dagger}\times\tilde d]^{(1)}$ being the boson angular momentum
operator and the gyro-magnetic factor $g_B=\mu_{2^+_1}/2$ 
given in terms of the magnetic moment $\mu_{2^+_1}$ of the $2^+_1$ state of the
even-even nucleus.  The corresponding experimental value has been  used for $\mu_{2^+_1}$. 
The fermionic part $\hat T_F^{(M1)}$ takes the form \cite{scholten1985}
\begin{equation}
 \hat T^{(M1)}_F=-\sum_{jj^{\prime}} g_{jj^{\prime}}
 \sqrt{\frac{j(j+1)(2j+1)}{3}}[a^{\dagger}_j\times\tilde a_{j^{\prime}}]^{(1)},
\end{equation}
with the coefficients $ g_{jj^{\prime}}$ given by
\begin{equation}
 g_{jj^{\prime}}
=\left\{\begin{array}{ll}
\frac{(2j-1)g_l + g_s}{2j} & (j=j^{\prime}=l+\frac{1}{2}) \\
  \frac{(2j+3)g_l - g_s}{2(j+1)} & (j=j^{\prime}=l-\frac{1}{2}) \\
	 (g_l-g_s)\sqrt{\frac{2l(l+1)}{j(j+1)(2j+1)(2l+1)}} &
	  (j^{\prime}=j-1; l=l^{\prime}) \\
	\end{array} \right.
\end{equation}
where $l$ represents the orbital angular momentum of the 
single-particle state. The fermion  $g_l$ and 
$g_s$ gyro-magnetic factors take the usual Schmidt values  $g_l=1.0$ 
$\mu^2_N$ and $g_s=5.58$ $\mu^2_N$ for the proton and $g_l=0$ and
$g_s=-3.82$ $\mu^2_N$ for the neutron. The $g_s$ value has been quenched by 30 \% for
both the protons and neutrons \cite{scholten1982,nomura2016odd}.

% ----------------------------------------------------------------------

\section{Results for the even-even core nuclei\label{sec:even-results}}

% ----------------------------------------------------------------------

% ----------------------------------------------------------------------
%                                            F i g u r e   5 
% ----------------------------------------------------------------------

\begin{figure*}[htb!]
\begin{center}
\includegraphics[width=\linewidth]{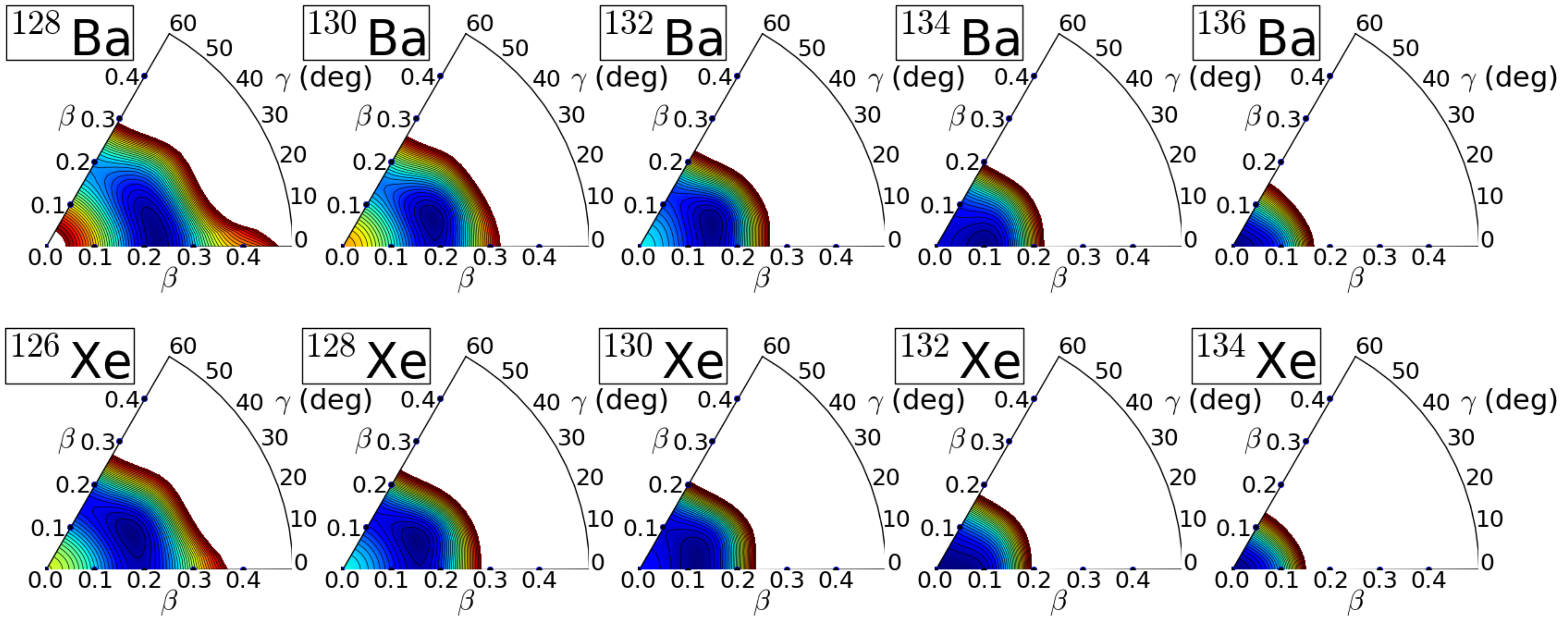}
\caption{(Color online) The Gogny-D1M energy surfaces 
 for the even-even nuclei $^{128-136}$Ba (upper panel) and $^{126-134}$Xe
 (lower panel) are plotted up to 3 MeV above the absolute 
 minimum. The difference between neighboring contours is 100 keV.}
\label{fig:pes-d1m}
\end{center}
\end{figure*}

% ----------------------------------------------------------------------
%                                            F i g u r e   6 
% ----------------------------------------------------------------------

\begin{figure*}[htb!]
\begin{center}
\includegraphics[width=\linewidth]{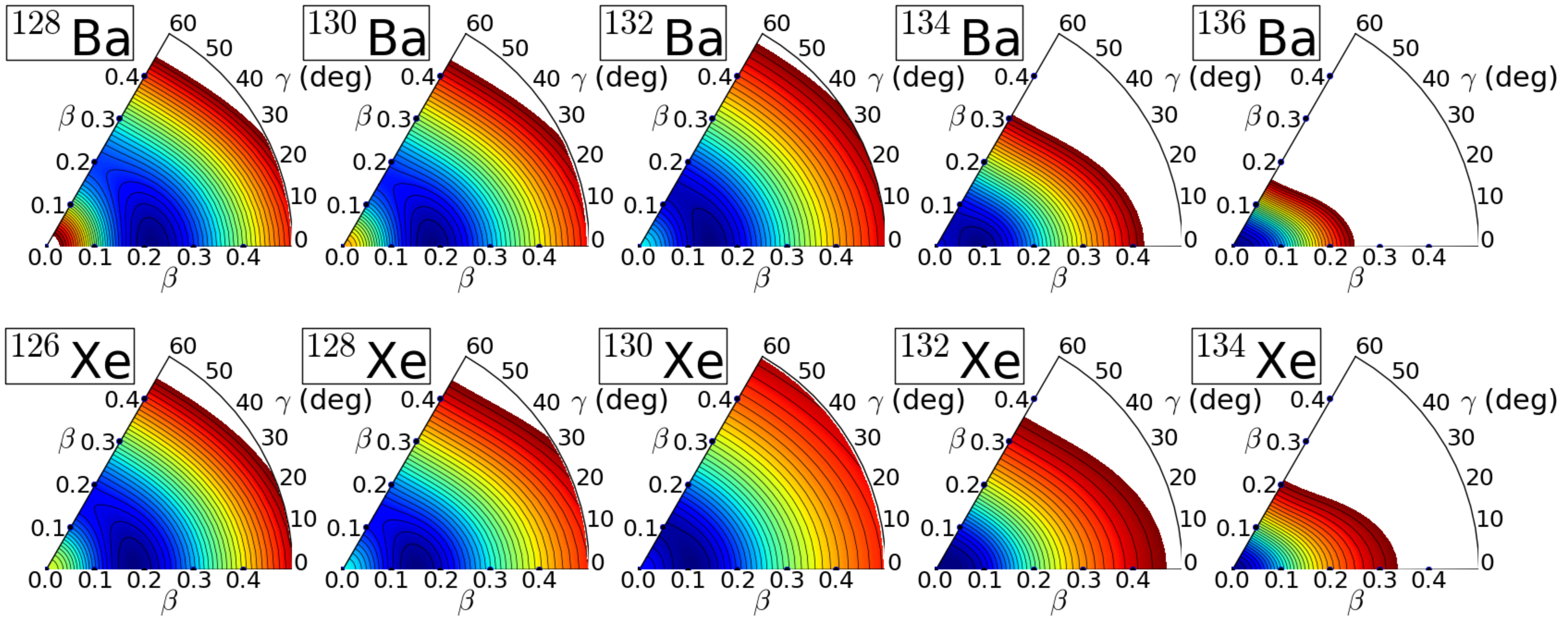}
\caption{(Color online) The same as in Fig.~\ref{fig:pes-d1m}, but for the mapped IBM energy surfaces.}
\label{fig:pes-mapped}
\end{center}
\end{figure*}

% ----------------------------------------------------------------------
%                                            F i g u r e    7
% ----------------------------------------------------------------------

\begin{figure}[htb!]
\begin{center}
\includegraphics[width=\linewidth]{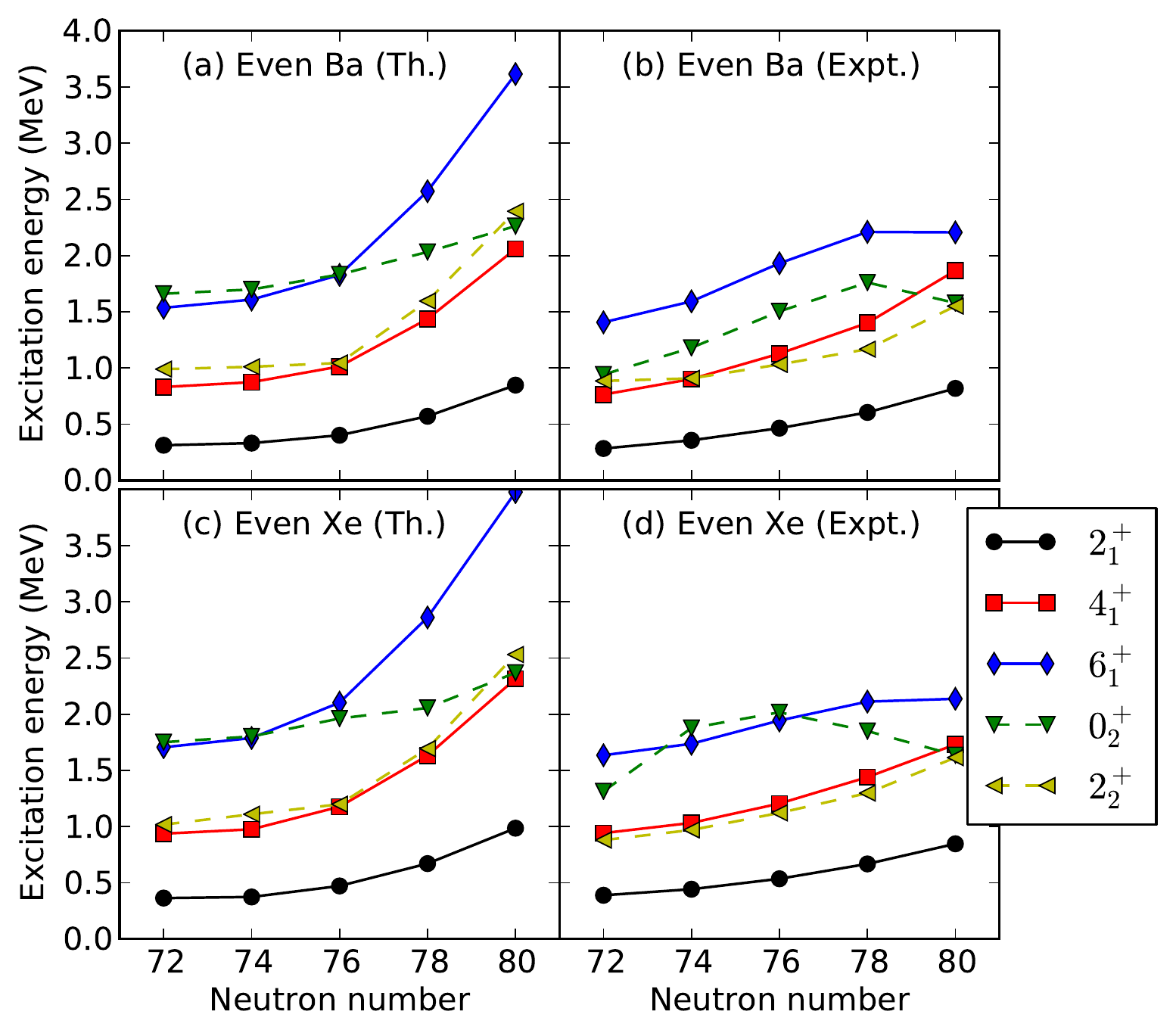}
\caption{(Color online) The low-energy excitation spectra obtained for $^{128-136}$Ba and
 $^{126-134}$Xe are plotted 
as functions of the neutron number $N$. Experimental data have been taken 
from Ref.~\cite{data}.}
\label{fig:even-level}
\end{center}
\end{figure}

In this section, we show that the IBM Hamiltonian, with the
parameters determined by mapping the Gogny-D1M energy surface onto the
expectation value of that Hamiltonian, provides a reasonable description of
the considered even-even core nuclei.  

The $(\beta,\gamma)$-deformation energy surfaces obtained for the nuclei
$^{128-136}$Ba and $^{126-134}$Xe within the constrained Gogny-D1M HFB 
method are shown in Fig.~\ref{fig:pes-d1m}. They suggest structural 
evolution from notably $\gamma$-soft ($^{128,130}$Ba and $^{126,128}$Xe) 
to nearly spherical ($^{136}$Ba and $^{134}$Xe) shapes. There are also  
transitional regions around $^{132,134}$Ba and $^{130,132}$Xe. In many 
cases, one observes a shallow triaxial minimum with $\gamma=0^{\circ} - 20^{\circ}$.
In general, the Gogny-D1M surfaces for the Xe nuclei look softer along the
$\gamma$-direction  than those for the Ba neighbors.

Of particular interest is $^{134}$Ba, identified as the first empirical 
evidence \cite{casten2000} of the E(5) critical-point symmetry 
\cite{iachello2000} of the second-order quantum phase transition between 
spherical vibrational U(5) and $\gamma$-soft O(6) dynamical symmetries. 
The E(5) model is derived from the five-dimensional collective Hamiltonian 
using a collective potential with an infinite square-well along $\beta$ 
and independent of the $\gamma$ deformation \cite{iachello2000}. As can 
be seen from Fig.~\ref{fig:pes-d1m}, the  Gogny-D1M energy surface for 
$^{134}$Ba looks almost flat for $0\le\beta\le 0.15$. It is also flat 
along the $\gamma$ direction. In fact, among all the studied even-even 
Ba and Xe nuclei, it is the one that best resembles the E(5)-potential. 
Note, that the Gogny-D1M energy surfaces depicted in Fig.~\ref{fig:pes-d1m} 
are, rather similar to those obtained within the relativistic framework 
based on the DD-PC1 \cite{DDPC1} EDF (see, Ref.~\cite{nomura2017odd-1}).

The mapped IBM surfaces are displayed in Fig.~\ref{fig:pes-mapped}. 
They exhibit a similar systematic trend, as functions of the
neutron number, as the original Gogny-D1M energy surfaces in
Fig.~\ref{fig:pes-d1m}. Despite the fact that only three parameters ($\epsilon_d$, $\kappa$ and
$\chi$) have been introduced to determine the IBM Hamiltonian, the mapped
surfaces  reproduce the most relevant feature of the Gogny-D1M
ones: $\gamma$-softness in the vicinity of the minimum.  
On the other hand, several discrepancies are also observed. At variance with
the mean-field ones, the IBM surfaces are generally flat  far from
the minimum, i.e., $\beta\ge 0.2$. 
One of the reasons for the discrepancy could be that we have paid
much attention to reproducing the topology of the Gogny-D1M energy surface in the 
neighborhood of the minimum as we expect that configurations around 
the minimum would be the  most relevant for the low-energy collective
dynamics. 
Another reason is that, like in previous  studies, the present
IBM framework only comprises  limited types (i.e., $J^\pi=0^+$ and $2^+$)
as well as number of valence nucleon pairs. Furthermore, it 
employs the simplified
form of the Hamiltonian Eq.~(\ref{eq:ibm}), whereas the HFB framework
contains more degrees of freedom and, as a consequence, provides energy
surfaces far richer in topology than the IBM ones.
Moreover,  in the case of lighter Ba and Xe isotopes, the IBM surfaces do not reproduce
the shallow triaxial minimum that is found in the Gogny-D1M ones, but
exhibit a minimum only at $\gamma=0^{\circ}$. 
This is because, as seen from the analytical expression of the energy
surface \cite{ginocchio1980}, any two-body IBM-1 Hamiltonian never gives rise to 
a triaxial minimum. To produce such a minimum, it is necessary to include a three-body
boson term \cite{vanisacker1981}. 
Such a  term is also important to better describe $\gamma$-band
levels, while it makes a marginal contribution to the ground-state band. 
For this reason, and for the sake of simplicity, we have not taken it 
into account in the considered IBM Hamiltonian.  

Even though the deformation energy surfaces provide useful information 
about shape phase transitions at the mean-field level, a more 
quantitative analysis requires the computation of spectroscopic 
properties. To this end, in Fig.~\ref{fig:even-level} we have plotted 
the low-energy spectra obtained for the considered even-even Ba and Xe 
nuclei as functions of the neutron number. They are compared with the 
available experimental data \cite{data}. Our results for the low-lying 
excitation spectra (panels (a) and (c) of Fig.~\ref{fig:even-level}) 
exhibit, a reasonable agreement with the experimental ones (see, panels 
(b) and (d)), exception made of the overestimation of the $6^+_1$ level 
at $N=80$. The spectra predicted for even-even Ba nuclei exhibit a 
steady increase with  neutron number (see, 
Fig.~\ref{fig:even-level}(a)). The near doublet ($4^+_1$, $2^+_2$), 
which is seen for $72\le N\le 78$, suggests $\gamma$-softness. At 
$N=80$, one notices the ($4^+_1$, $2^+_2$,$0^+_2$) triplet, 
characteristic of a vibrational nucleus.  As seen from 
Fig.~\ref{fig:even-level}(c), a similar overall trend  has been 
obtained for  even-even Xe nuclei. Though the spectra obtained in 
Ref.~\cite{nomura2017odd-1} are similar, they are more stretched than 
those obtained in the present work. 

% ----------------------------------------------------------------------

\section{Results for the odd-$A$ nuclei\label{sec:odd-results}}

% ----------------------------------------------------------------------

% ----------------------------------------------------------------------
%                                            F i g u r e     8 
% ----------------------------------------------------------------------

\begin{figure}[htb!]
\begin{center}
\includegraphics[width=\linewidth]{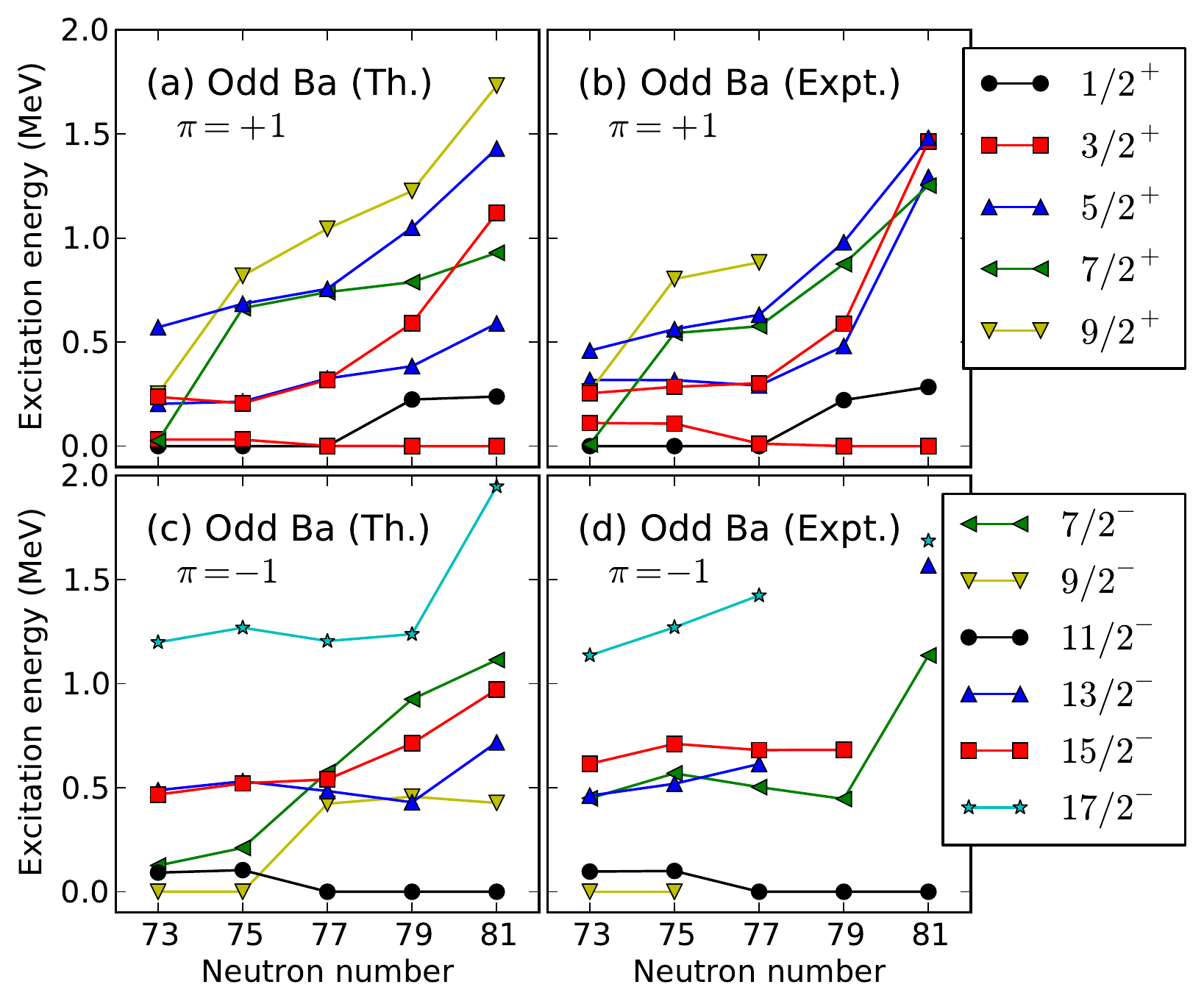}
\caption{(Color online) The low-energy positive- ($\pi=+1$) and
negative-parity ($\pi=-1$) excitation spectra in the odd-mass isotopes $^{129-137}$Ba 
are plotted as functions of the neutron number $N$. Experimental 
data have been taken from Ref.~\cite{data}.} 
\label{fig:ba-level}
\end{center}
\end{figure}

% ----------------------------------------------------------------------
%                                            F i g u r e     9
% ----------------------------------------------------------------------

\begin{figure}[htb!]
\begin{center}
\includegraphics[width=\linewidth]{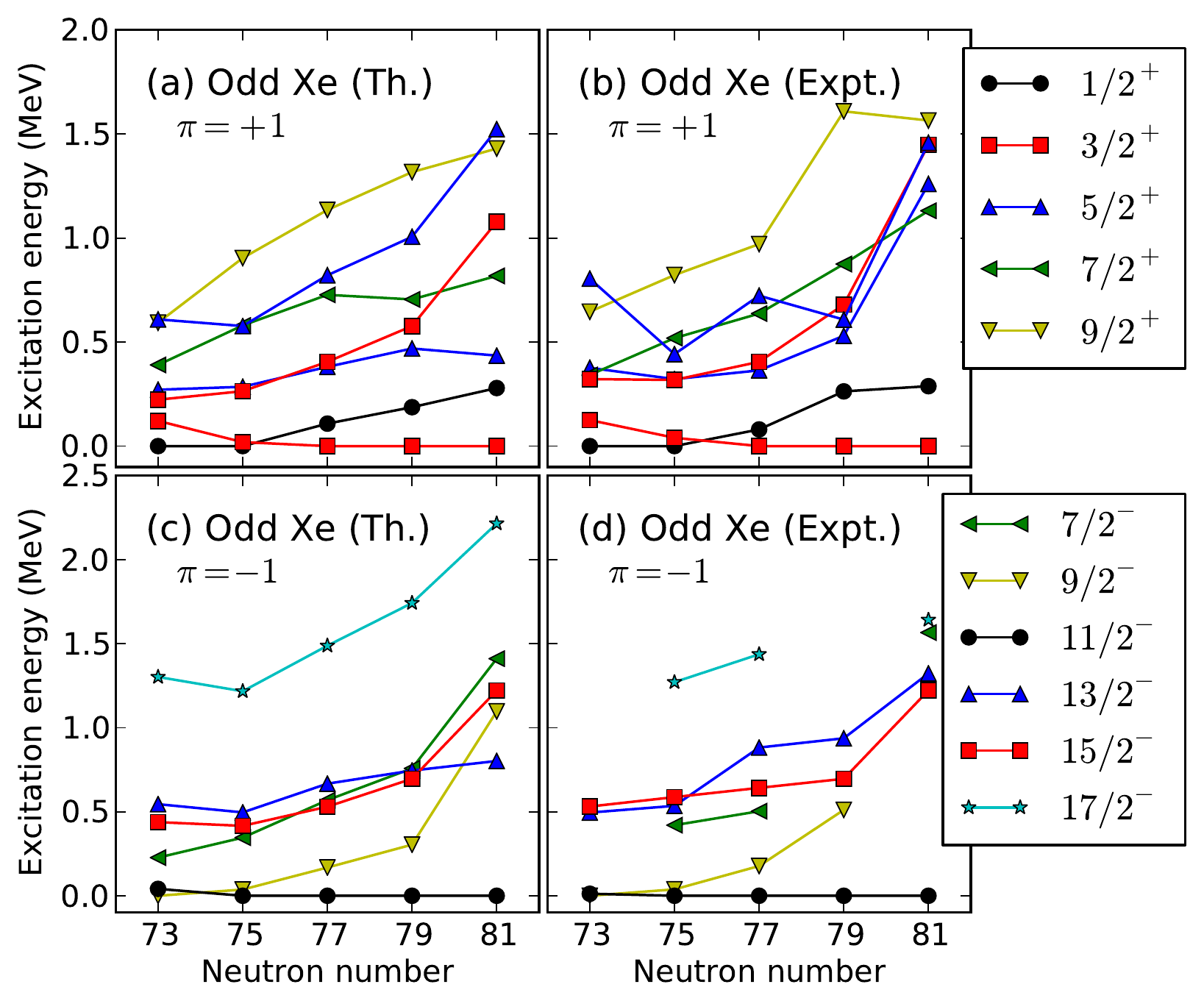}
\caption{(Color online) The same as in Fig.~\ref{fig:ba-level}, but for the
odd-mass Xe isotopes.}
\label{fig:xe-level}
\end{center}
\end{figure}

% ----------------------------------------------------------------------
%                                            F i g u r e   1 0
% ----------------------------------------------------------------------

\begin{figure}[htb!]
\begin{center}
\includegraphics[width=\linewidth]{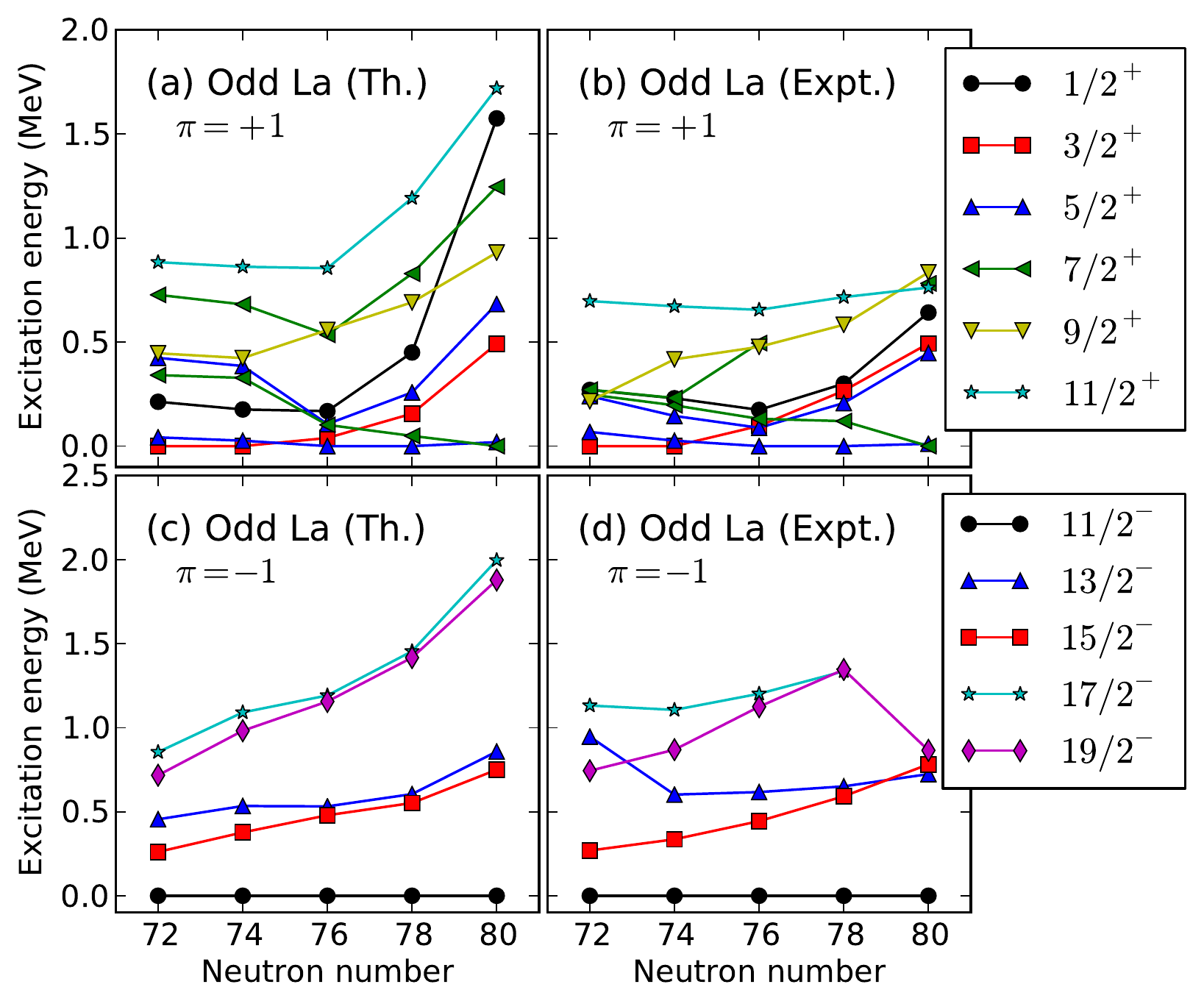}
\caption{(Color online) The same as in Fig.~\ref{fig:ba-level}, but for the
odd-mass La isotopes.}
\label{fig:la-level}
\end{center}
\end{figure}

% ----------------------------------------------------------------------
%                                            F i g u r e   1 1
% ----------------------------------------------------------------------

\begin{figure}[htb!]
\begin{center}
\includegraphics[width=\linewidth]{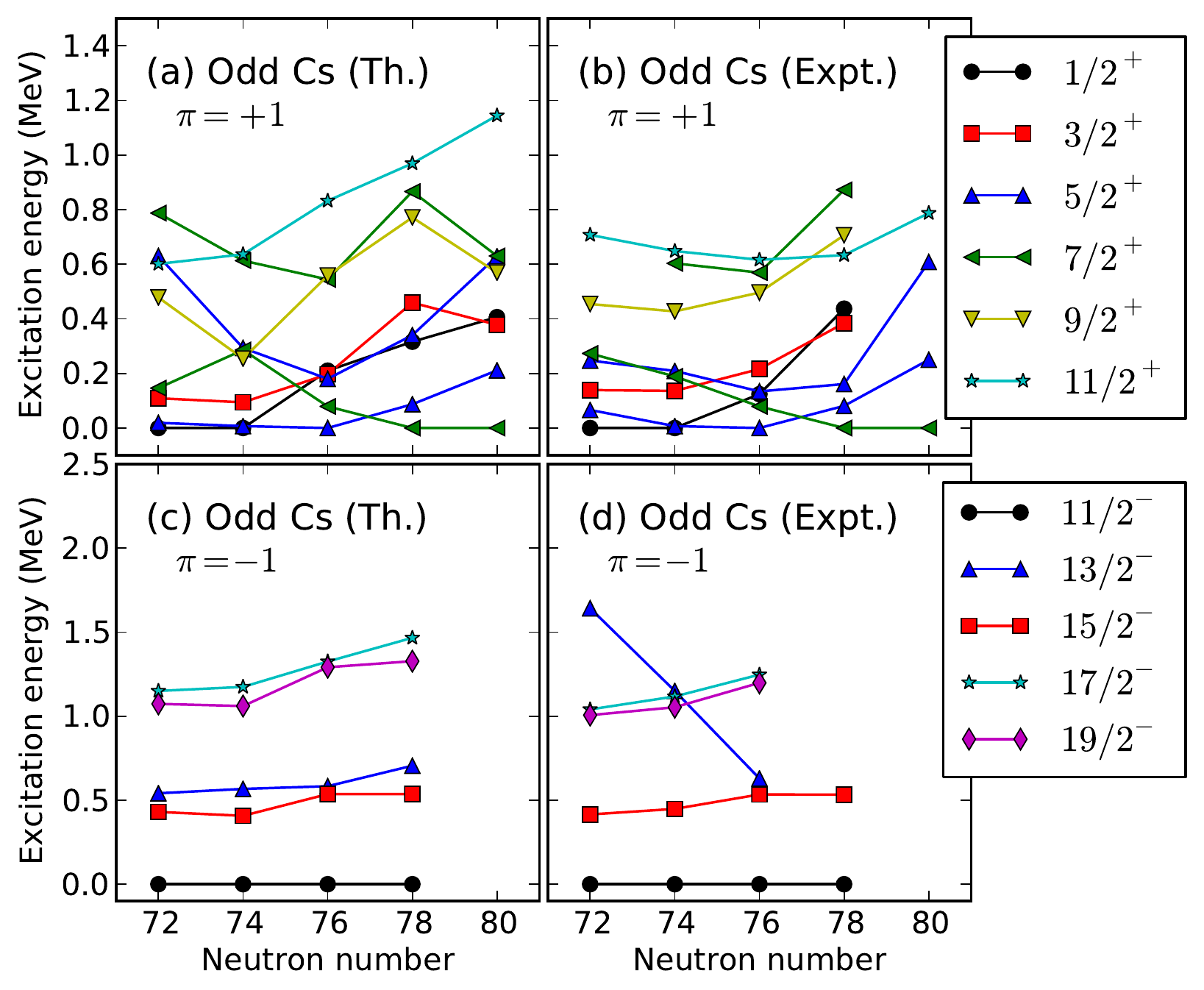}
\caption{(Color online) The same as in Fig.~\ref{fig:ba-level}, but for the
odd-mass Cs isotopes.}
\label{fig:cs-level}
\end{center}
\end{figure}

In this section, we turn our attention to the
odd-mass  nuclei $^{129-137}$Ba, $^{127-135}$Xe,
$^{129-137}$La and $^{127-135}$Cs.
We discuss the overall  trend of the computed 
excitation spectra (Sec.~\ref{sec:spectra}) as well as  electromagnetic
transitions (Sec.~\ref{sec:em}) in each isotopic chain as
functions of the nucleon number. We then present more detailed spectroscopic results 
for some selected nuclei (Sec.~\ref{sec:detail}).

\subsection{Systematics of low-energy spectra\label{sec:spectra}}

In Figs.~\ref{fig:ba-level} to \ref{fig:cs-level}, the low-energy
positive- and negative-parity excitation spectra obtained for the considered
odd-mass Ba, Xe, La and Cs nuclei are depicted  as functions of the
neutron number. 
They have been obtained via the diagonalization of the IBFM Hamiltonian with 
the parameters shown in Tables \ref{tab:paraB} and
\ref{tab:monopole} (see also,  Figs.~\ref{fig:spe}-\ref{fig:para-exc}). 
For each isotopic chain, the predicted spectra are compared with the
available experimental values in Ref.~\cite{data}.
Those spectra are in reasonable good agreement with the 
experiment.

One signature of structural evolution in odd-mass nuclei is given by a change 
in the ground state spin. For example, the 
positive-parity levels corresponding to Ba isotopes in 
Figs.~\ref{fig:ba-level}(a) and \ref{fig:ba-level}(b), display a ground state spin change
from $J^{\pi}={1/2}^+$ to ${3/2}^+$ at $N=79$. Both 
theoretically and experimentally,  the ${7/2}^+_1$ and ${9/2}^+_1$ levels at
$N=73$ are significantly lower in energy with respect to the neighboring
isotope $^{131}$Ba. This is the consequence of the fact that 
the $1g_{7/2}$ single-particle
orbital becomes lower and closer in energy to the $3s_{1/2}$ and
$2d_{3/2}$ orbitals at $N=73$. The 
${7/2}^+_1$ and ${9/2}^+_1$ states obtained in the calculations 
are almost purely (96 \% and 97 \%, respectively)
composed of the $1g_{7/2}$ configuration.
Furthermore, the negative-parity spectra in Figs.~\ref{fig:ba-level}(c) and \ref{fig:ba-level}(d) 
suggest a change in the ground-state spin from $N=75$ to 77. 
Similar results  are obtained for
Xe isotopes (see, Fig.~\ref{fig:xe-level}). 
In the case of  $^{129-137}$La
(Figs.~\ref{fig:la-level}(a) and \ref{fig:la-level}(c)) and $^{127-135}$Cs
(Figs.~\ref{fig:cs-level}(a) and \ref{fig:cs-level}(c)), a reasonable agreement
with the experimental data is also observed. Both the 
theoretical and experimental systematics of the 
positive-parity states for
La (Figs.~\ref{fig:la-level}(a) and \ref{fig:la-level}(b)) and Cs
(Figs.~\ref{fig:cs-level}(a) and \ref{fig:cs-level}(b)) nuclei, show 
many levels with very low excitation energy (below $E_x\approx 0.2$
MeV) at $N\approx 76$ . This reflects structural changes 
around this neutron number.
Indeed, the Gogny-D1M energy surfaces for the corresponding even-even Ba
and Xe nuclei around $N\approx 76$ suggest that it corresponds to the
transition point between nearly spherical and
$\gamma$-soft shapes (see, Fig.~\ref{fig:pes-d1m}).

% ----------------------------------------------------------------------
%                                            F i g u r e   1 2
% ----------------------------------------------------------------------

\begin{figure}[htb!]
\begin{center}
\includegraphics[width=\linewidth]{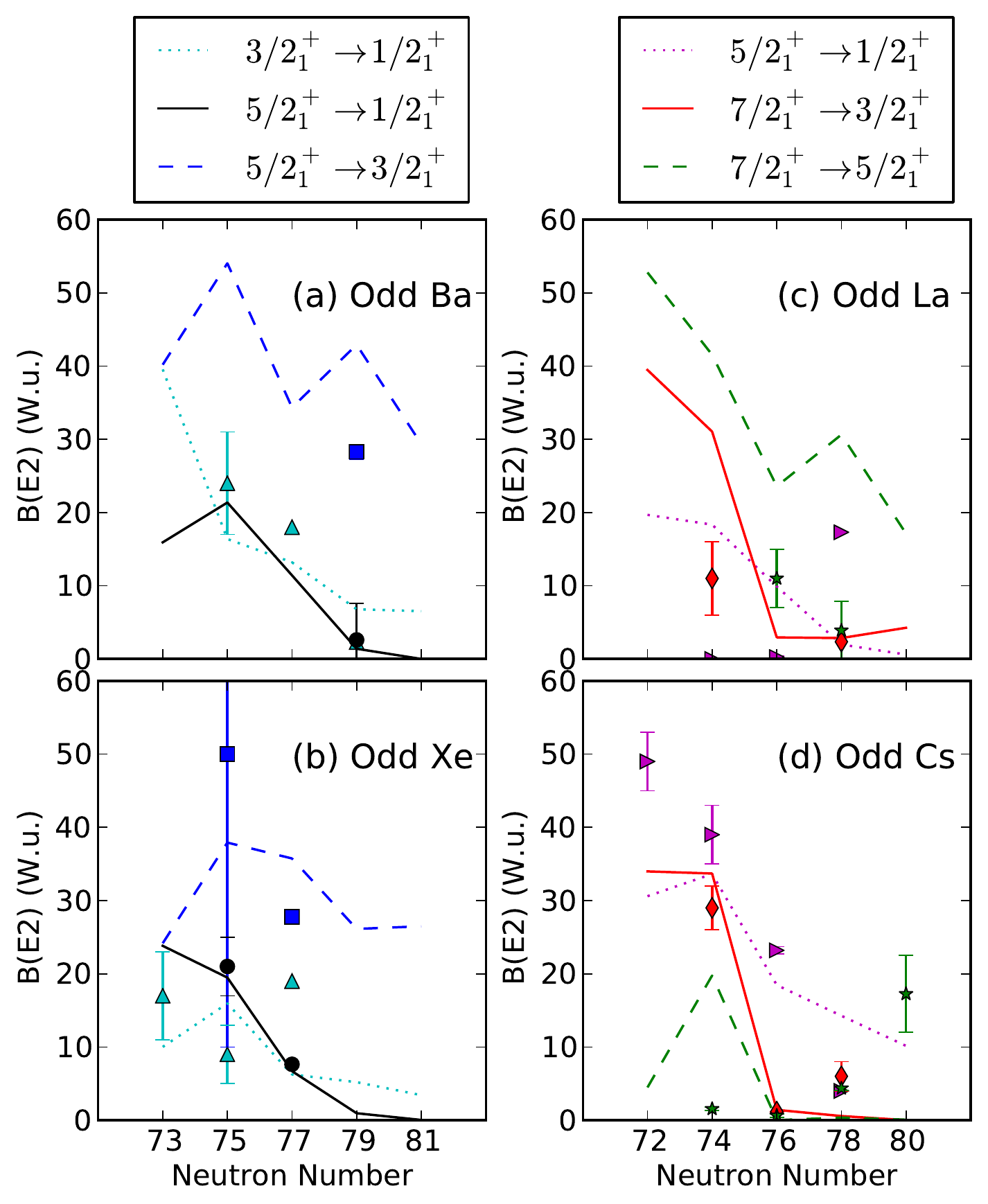}
\caption{(Color online) The transition probabilities 
$B(E2; {3/2}^+_1\rightarrow {1/2}^+_1)$, $B(E2; {5/2}^+_1\rightarrow
 {1/2}^+_1)$ and $B(E2; {5/2}^+_1\rightarrow {3/2}^+_1)$ 
 obtained for the odd-mass nuclei 
 $^{129-137}$Ba (a)  and $^{127-135}$Xe (b), and the
 transition probabilities
 $B(E2; {5/2}^+_1\rightarrow {1/2}^+_1)$, $B(E2; {7/2}^+_1\rightarrow  {3/2}^+_1)$ and
 $B(E2; {7/2}^+_1\rightarrow {5/2}^+_1)$ 
 obtained for the odd-mass nuclei
 $^{129-137}$La (c) and $^{127-135}$Cs (d) are plotted as functions of the neutron
 number. The experimental $B(E2;{3/2}^+_1\rightarrow {1/2}^+_1)$ (filled circles),
 $B(E2;{5/2}^+_1\rightarrow {1/2}^+_1)$ (squares), 
 $B(E2;{5/2}^+_1\rightarrow {3/2}^+_1)$ (triangles),
 $B(E2;{7/2}^+_1\rightarrow {3/2}^+_1)$ (diamonds)
 and $B(E2;{7/2}^+_1\rightarrow {5/2}^+_1)$ (asterisks) values are 
 taken from
 Ref.~\cite{data}. 
}
\label{fig:e2}
\end{center}
\end{figure}

% ----------------------------------------------------------------------
%                                            F i g u r e   1 3
% ----------------------------------------------------------------------

\begin{figure}[htb!]
\begin{center}
\includegraphics[width=\linewidth]{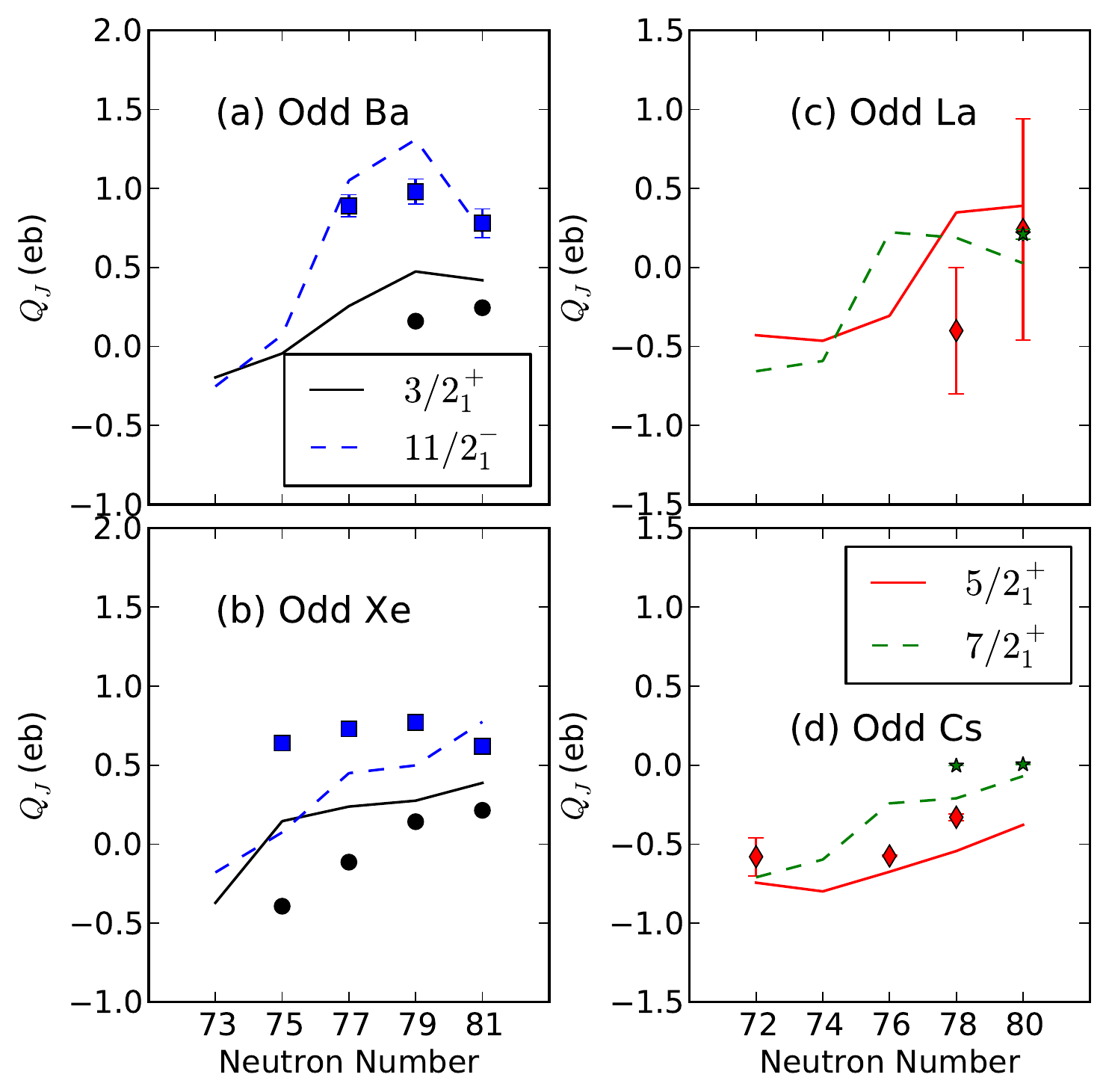}
\caption{(Color online) The spectroscopic quadrupole moments
 $Q_{{3/2}^+_1}$ and $Q_{{11/2}^-_1}$ 
 obtained for $^{129-137}$Ba (a) 
 and $^{127-135}$Xe (b) and 
 the spectroscopic quadrupole moments
 $Q_{{5/2}^+_1}$ and $Q_{{7/2}^+_1}$ 
 obtained for
 $^{129-137}$La (c) and $^{127-135}$Cs (d)
 are plotted as functions of
 the neutron number. 
 The experimental $Q_{{3/2}^+_1}$ (filled circles),
 $Q_{{11/2}^-_1}$ (squares),
 $Q_{{5/2}^+_1}$  (diamonds) and 
 $Q_{{7/2}^+_1}$ (asterisks) values are taken from 
 Ref.~\cite{stone2005}.}
\label{fig:qmom}
\end{center}
\end{figure}

% ----------------------------------------------------------------------
%                                            F i g u r e   1 4
% ----------------------------------------------------------------------

\begin{figure}[htb!]
\begin{center}
\includegraphics[width=\linewidth]{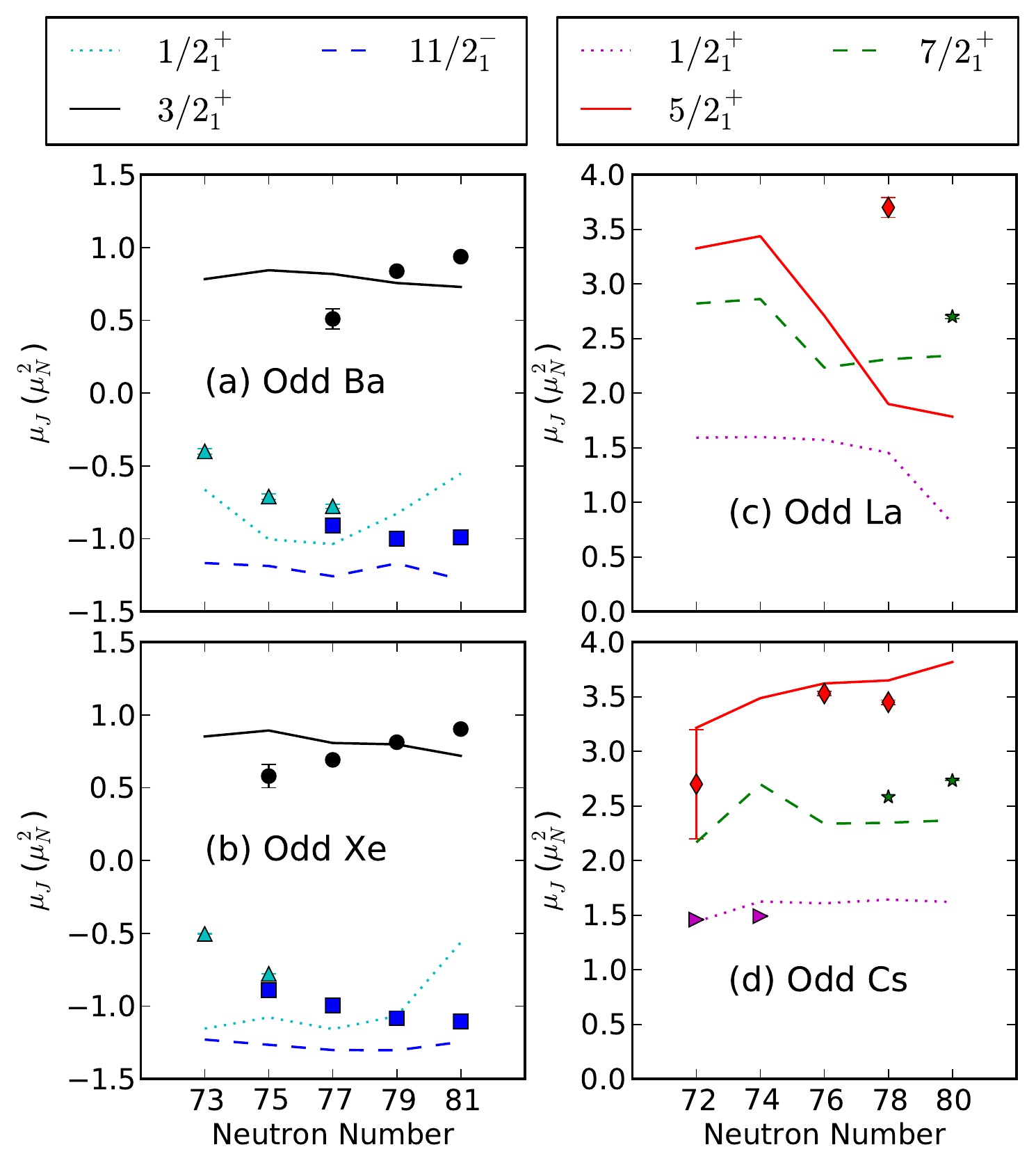}
\caption{(Color online) The magnetic moments $\mu_{{1/2}^+_1}$,
 $\mu_{{3/2}^+_1}$ and $\mu_{{11/2}^-_1}$ obtained for  $^{129-137}$Ba
 (a) 
 and $^{127-135}$Xe (b) and 
 the magnetic moments
 $\mu_{{1/2}^+_1}$, $\mu_{{5/2}^+_1}$ and $\mu_{{7/2}^+_1}$ 
 obtained for $^{129-137}$La (c) and $^{127-135}$Cs (d)
 are plotted as functions of the neutron number. 
 The experimental $\mu_{{1/2}^+_1}$ (filled triangles),
 $\mu_{{3/2}^+_1}$ (circles),
 $\mu_{{11/2}^-_1}$ (squares),
 $\mu_{{5/2}^+_1}$ (diamonds)
 and $\mu_{{7/2}^+_1}$ (asterisks)
 are taken from Ref.~\cite{stone2005}.}
\label{fig:mmom}
\end{center}
\end{figure}

\subsection{Systematics of electromagnetic properties\label{sec:em}}

The $B(E2)$ transition rates between 
low-lying positive-parity states where experimental data are available
are shown in Fig.~\ref{fig:e2}. For each of the considered isotopic 
chains, the $B(E2)$ values increase as the valence nucleon number 
increases reflecting the
development of quadrupole collectivity in the corresponding even-even
systems. The computed $B(E2)$ values appear to be in a reasonable
agreement with the experiment. However, the $B(E2;
{5/2}^+_1\rightarrow {1/2}^+_1)$ in odd-mass La nuclei (panel
(c) of Fig.~\ref{fig:e2}), for example, decreases (dotted line)
while experimentally (triangle) it 
increases from $N=74$ to 78. The analysis of the compositions of the
wave functions reveals that the ${5/2}^+_1$ and
${1/2}^+_1$ states are rather 
similar in structure for the odd-$A$ La isotopes with $N\leq 76$ (both are mainly
composed of the $2d_{5/2}$ single-particle configuration) leading to
large E2 matrix elements. On the other hand, for La isotopes with
$N\geq 78$ the computed wave functions 
have rather different compositions, i.e., the dominant components of 
the ${5/2}^+_1$ and ${1/2}^+_1$ states are the $1g_{7/2}$
and $2d_{5/2}$ configurations, respectively. 

In Fig.~\ref{fig:qmom} the predicted spectroscopic quadrupole $Q_J$ 
moments corresponding to the ${3/2}^+_1$ and ${11/2}^-_1$ states of 
the odd-mass Ba (panel (a)) and Xe (panel (b)) nuclei are plotted as 
functions of the neutron number. The spectroscopic quadrupole moments  
associated with the ${5/2}^+_1$ and ${7/2}^+_1$ states of the odd-mass La 
(panel (c)) and Cs (panel (d)) nuclei are also shown in the plot. 
The predicted $Q_J$ values agree reasonably well with the available 
experimental data \cite{stone2005} exception made of the lighter 
($N=75$ and 77) Xe isotopes. 

The magnetic moments $\mu_{{1/2}^+_1}$, $\mu_{{3/2}^+_1}$ and
$\mu_{{11/2}^-_1}$ obtained for the 
odd-$A$ Ba (panel (a)) and Xe (panel (b)) nuclei are plotted in Fig.~\ref{fig:mmom}.
The $\mu_{{1/2}^+_1}$, $\mu_{{5/2}^+_1}$ and $\mu_{{7/2}^+_1}$ moments
obtained for the
odd-mass La (panel (c)) and Cs (panel (d)) nuclei are also included in the plot.
The comparison with 
the available experimental data \cite{stone2005} reveals a reasonable agreement
both in magnitude and sign.

% ----------------------------------------------------------------------
%                                            F i g u r e   1 5
% ----------------------------------------------------------------------

\begin{figure}[htb!]
\begin{center}
\includegraphics[width=\linewidth]{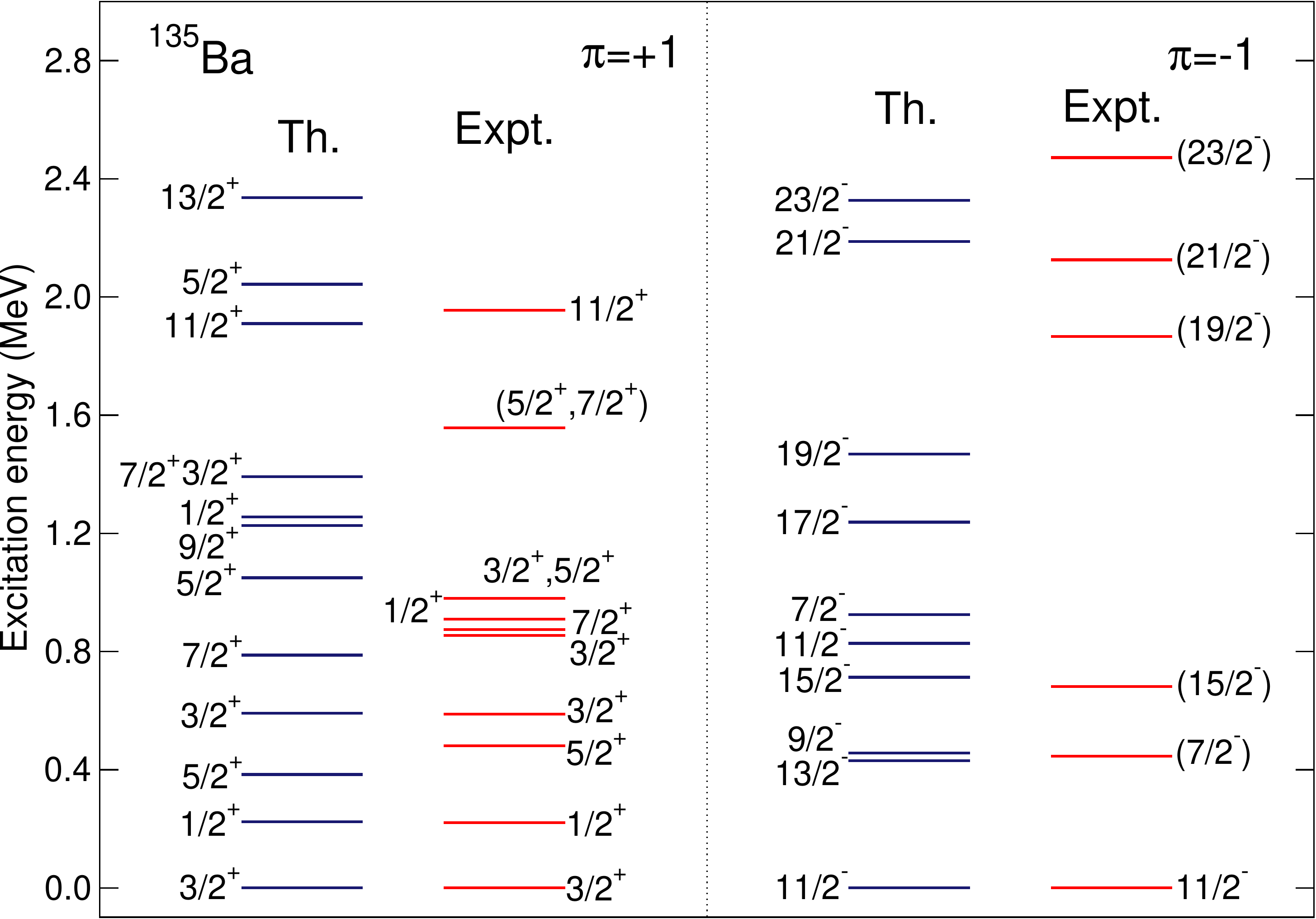}
\caption{(Color online) Detailed level scheme for the low-energy
 positive- and negative-parity states in  $^{135}$Ba. Experimental data are
  from Ref.~\cite{data}.}
\label{fig:135ba}
\end{center}
\end{figure}

% ----------------------------------------------------------------------
%                                                      T a b l e   I I I
% ----------------------------------------------------------------------

\begin{table}[htb]
\caption{\label{tab:135ba}% 
Comparison between the theoretical and experimental $B(E2)$ and $B(M1)$
values in $^{135}$Ba. Experimental  data are taken from Ref.~\cite{data}.}
\begin{center}
\begin{tabular}{p{2.0cm}cccc}
\hline\hline
\multirow{2}{*}{} & \multicolumn{2}{c}{$B(E2)$ (W.u.)} &
 \multicolumn{2}{c}{$B(M1)$ (W.u.)} \\
\cline{2-3} 
\cline{4-5}
          & Th.         & Expt.    & Th.         & Expt.     \\
\hline
${1/2}^+_1\rightarrow {3/2}^+_1$ & 14 & 4.6(2) & 0.0010 & 0.0025(11) \\
${1/2}^+_2\rightarrow {3/2}^+_1$ & 7.2 & 11.7(10) &  &  \\
${3/2}^+_2\rightarrow {3/2}^+_1$ & 7.5 & 18.0(10) &  &  \\
${3/2}^+_3\rightarrow {3/2}^+_1$ & 1.1 & 7.0(10) &  &  \\
${5/2}^+_1\rightarrow {1/2}^+_1$ & 1.4 & 2.6(5) &  &  \\
${5/2}^+_1\rightarrow {3/2}^+_1$ & 43 & 28.3(10) & 3.7$\times 10^{-5}$ & 0.0042(20) \\
${7/2}^+_1\rightarrow {3/2}^+_1$ & 29 & 19.9(8) &  &  \\
${7/2}^+_1\rightarrow {5/2}^+_1$ & 25 & 12.8(12) & 0.0015 & 0.0032(3) \\
\hline\hline
\end{tabular}
\end{center}
\end{table}

% ----------------------------------------------------------------------
%                                            F i g u r e   1 6
% ----------------------------------------------------------------------

\begin{figure}[htb!]
\begin{center}
\includegraphics[width=\linewidth]{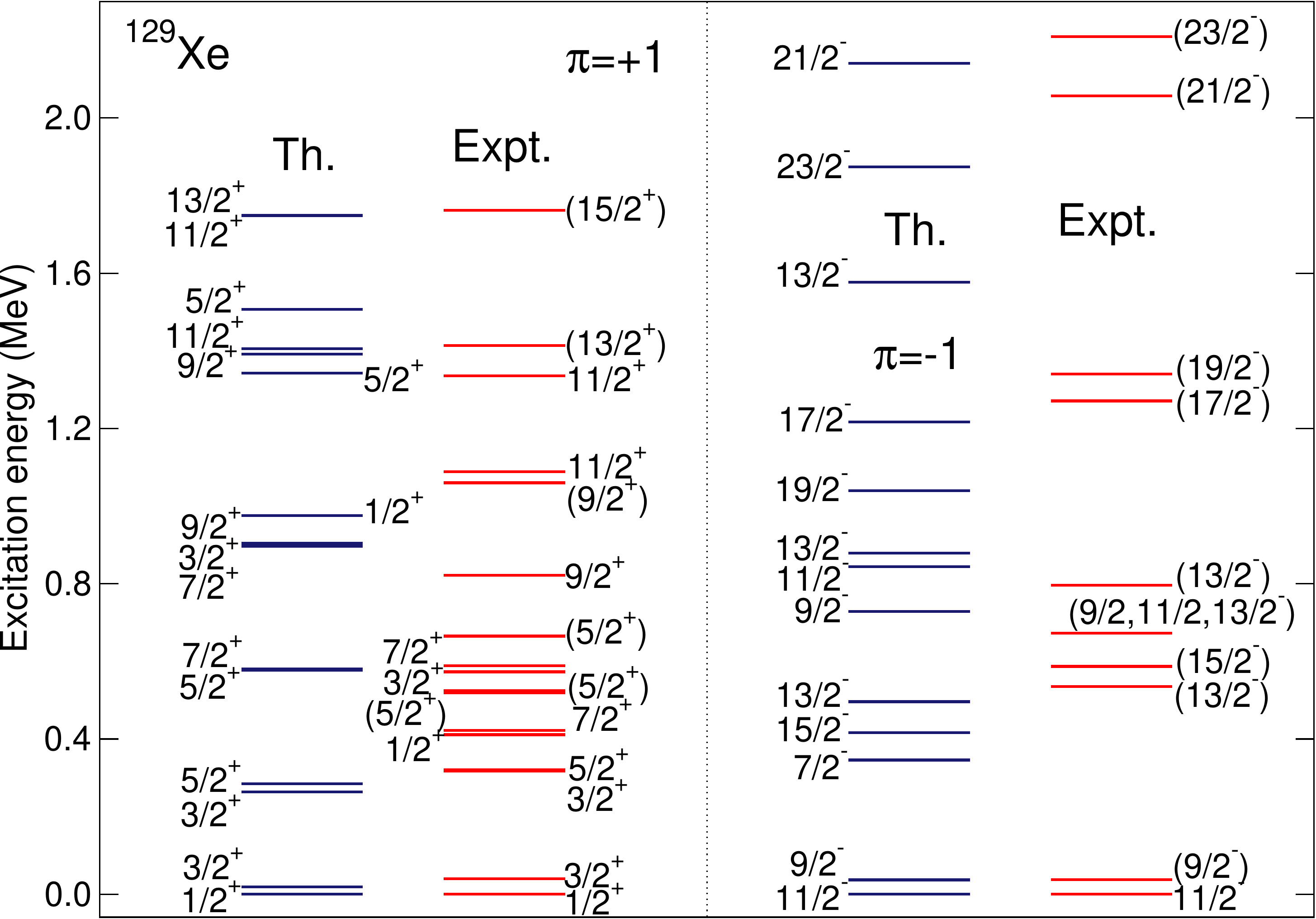}
\caption{(Color online) The same as in Fig.~\ref{fig:135ba}, but for $^{129}$Xe.}
\label{fig:129xe}
\end{center}
\end{figure}

% ----------------------------------------------------------------------
%                                                      T a b l e   I V
% ----------------------------------------------------------------------

\begin{table}[htb]
\caption{\label{tab:129xe}%
The same as in Table~\ref{tab:135ba}, but for $^{129}$Xe. }
\begin{center}
\begin{tabular}{p{2.0cm}cccc}
\hline\hline
\multirow{2}{*}{} & \multicolumn{2}{c}{$B(E2)$ (W.u.)} &
 \multicolumn{2}{c}{$B(M1)$ (W.u.)} \\
\cline{2-3} 
\cline{4-5}
          & Th.         & Expt.    & Th.         & Expt.     \\
\hline
${1/2}^+_{2}\rightarrow {1/2}^+_{1}$ &  &  & 0.011 & 0.0016(5) \\
${1/2}^+_{2}\rightarrow {3/2}^+_{1}$ & 14 & 6.7(23) & 0.031 &
 0.0039(13) \\
${1/2}^+_{2}\rightarrow {3/2}^+_{2}$ &  &  & 0.011 & 0.0015(5) \\
${1/2}^+_{2}\rightarrow {5/2}^+_{1}$ & 1.6 & 1.4(6) &  &  \\
${3/2}^+_{1}\rightarrow {1/2}^+_{1}$ & 16 & 9(4) & 0.035 & 0.0281(7) \\
${3/2}^+_{2}\rightarrow {1/2}^+_{1}$ & 25 & 23$^{+25}_{-23}$ &  &
  \\
${3/2}^+_{2}\rightarrow {3/2}^+_{1}$ & 36 & 17$^{+27}_{-17}$ & 0.0078 & 0.003$^{+4}_{-3}$ \\
${3/2}^+_{3}\rightarrow {1/2}^+_{1}$ & 7.8 & $>$0.2 & 0.017 & $>$0.0001 \\
${3/2}^+_{3}\rightarrow {1/2}^+_{2}$ & 2.2 & $>$5.9 & 0.0048 & $>$0.0026 \\
${3/2}^+_{3}\rightarrow {3/2}^+_{1}$ & 9.7 & $>$1.6 & 0.0083 & $>$0.00071 \\
${3/2}^+_{3}\rightarrow {3/2}^+_{2}$ & 0.80 & $>$3.4 & 0.015 & $>$0.00037 \\
${3/2}^+_{3}\rightarrow {5/2}^+_{1}$ & 3.8 & $>$4.6 & 0.0083 & $>$0.0005 \\
${5/2}^+_{1}\rightarrow {1/2}^+_{1}$ & 19 & 21(4) &  &  \\
${5/2}^+_{1}\rightarrow {3/2}^+_{1}$ & 38 & 5${\times}10^1$(4) &
 0.027 & 0.011(5) \\
${5/2}^+_{4}\rightarrow {1/2}^+_{1}$ & 0.00014 & 15.4(19) &  &  \\
\hline\hline
\end{tabular}
\end{center}
\end{table}

% ----------------------------------------------------------------------
%                                            F i g u r e   1 7
% ----------------------------------------------------------------------

\begin{figure}[htb!]
\begin{center}
\includegraphics[width=\linewidth]{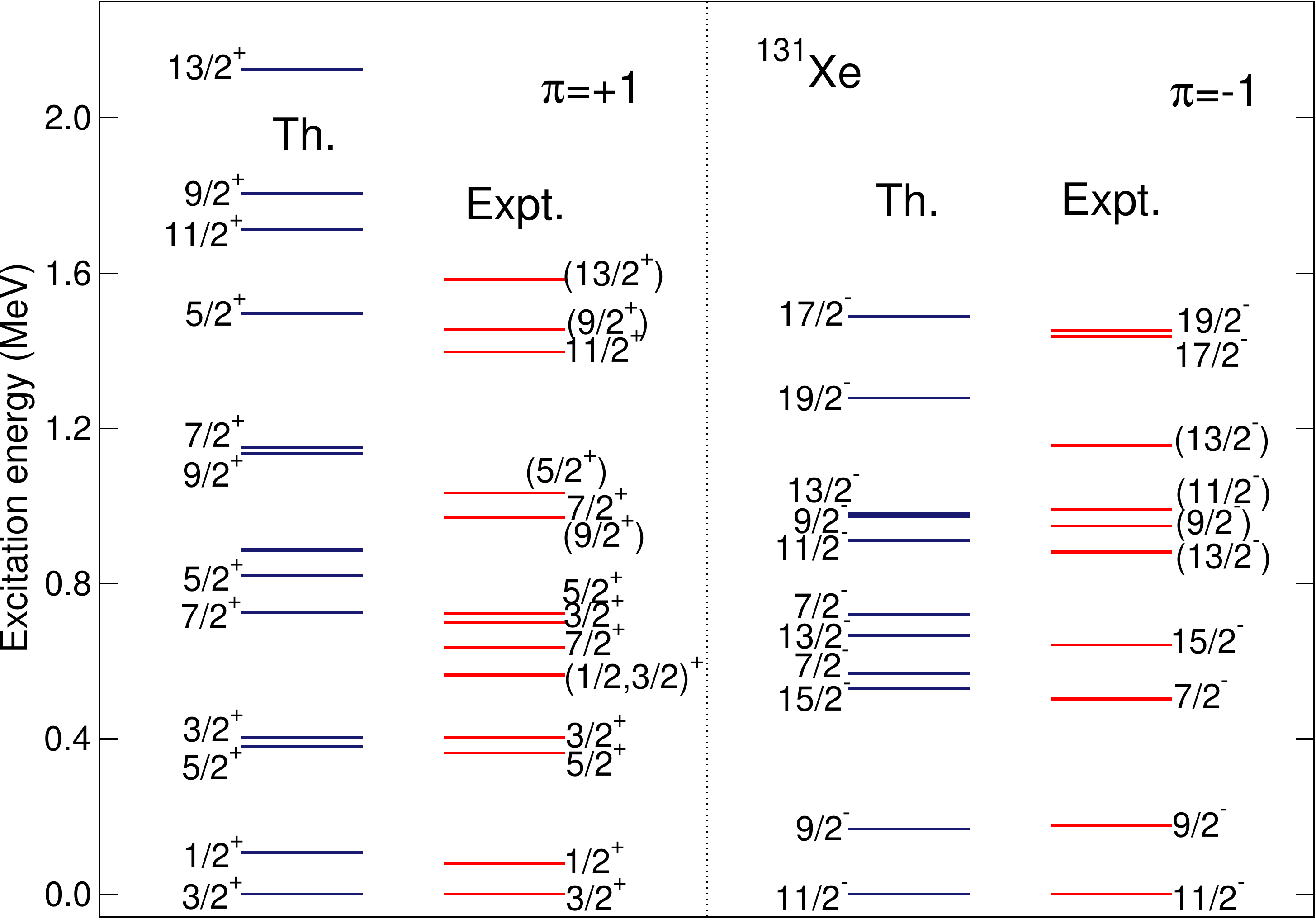}
\caption{(Color online) The same as in Fig.~\ref{fig:135ba}, but for $^{131}$Xe.}
\label{fig:131xe}
\end{center}
\end{figure}

% ----------------------------------------------------------------------
%                                                      T a b l e   V
% ----------------------------------------------------------------------

\begin{table}[htb]
\caption{\label{tab:131xe}
The same as in Table~\ref{tab:135ba}, but for $^{131}$Xe. }
\begin{center}
\begin{tabular}{p{2.2cm}cccc}
\hline\hline
\multirow{2}{*}{} & \multicolumn{2}{c}{$B(E2)$ (W.u.)} &
 \multicolumn{2}{c}{$B(M1)$ (W.u.)} \\
\cline{2-3} 
\cline{4-5}
          & Th.         & Expt.    & Th.         & Expt.     \\
\hline
${1/2}^+_{1}\rightarrow {3/2}^+_{1}$ & 13 & $<$37 & 0.00021 & $>$0.032 \\
${5/2}^+_{1}\rightarrow {1/2}^+_{1}$ & 6.7 & 7.64(24) & & \\
${5/2}^+_{1}\rightarrow {3/2}^+_{1}$ & 36 & 27.8(9) & 0.0013
 & 0.00030(3) \\
${3/2}^+_{2}\rightarrow {1/2}^+_{1}$ & 18 & 24$^{+26}_{-24}$ &
 0.0015 & 0.006(5) \\
${3/2}^+_{2}\rightarrow {1/2}^+_{1}$ & 9.8 & 3.$\times 10^1(3)$ &
 0.00019 & 0.006(6) \\
${3/2}^+_{3}\rightarrow {3/2}^+_{1}$ & 7.2 & 10(6) & 0.95 &
 0.014(4) \\
${1/2}^+_{2}\rightarrow {3/2}^+_{3}$ & 1.5 & 10(6) & 3.1 &
 0.014(4) \\
${7/2}^+_{1}\rightarrow {3/2}^+_{2}$ & 0.79 & 1.52(25) & & \\
${7/2}^+_{1}\rightarrow {5/2}^+_{1}$ & 6.7 & 1.6(13) & 0.0041
 & 0.00124(18) \\
${7/2}^+_{1}\rightarrow {3/2}^+_{1}$ & 35 & 22.2(19) & & \\
${5/2}^+_{2}\rightarrow {3/2}^+_{2}$ & 5.7 & 4$^{+6}_{-4}$ &
 0.14 & 0.047(5) \\
${5/2}^+_{2}\rightarrow {1/2}^+_{1}$ & 30 & 25.7(25) & & \\
${5/2}^+_{2}\rightarrow {3/2}^+_{1}$ & 5.5 & 4.8(5) & 0.035 & 0.090(9) \\
${7/2}^+_{2}\rightarrow {3/2}^+_{1}$ & 0.0018 & 1.6(7) & & \\
${9/2}^-_{1}\rightarrow {11/2}^-_{1}$ & 42 & 39(10) & 0.094 &
 0.00010(4) \\
${7/2}^-_{1}\rightarrow {9/2}^-_{1}$ & 0.050 & 0.17(6) &
 0.00010 & 0.000511(9) \\
${7/2}^-_{1}\rightarrow {11/2}^-_{1}$ & 35 & 0.494532(20) & & \\
\hline\hline
\end{tabular}
\end{center}
\end{table}

% ----------------------------------------------------------------------
%                                            F i g u r e   1 8
% ----------------------------------------------------------------------

\begin{figure}[htb!]
\begin{center}
\includegraphics[width=\linewidth]{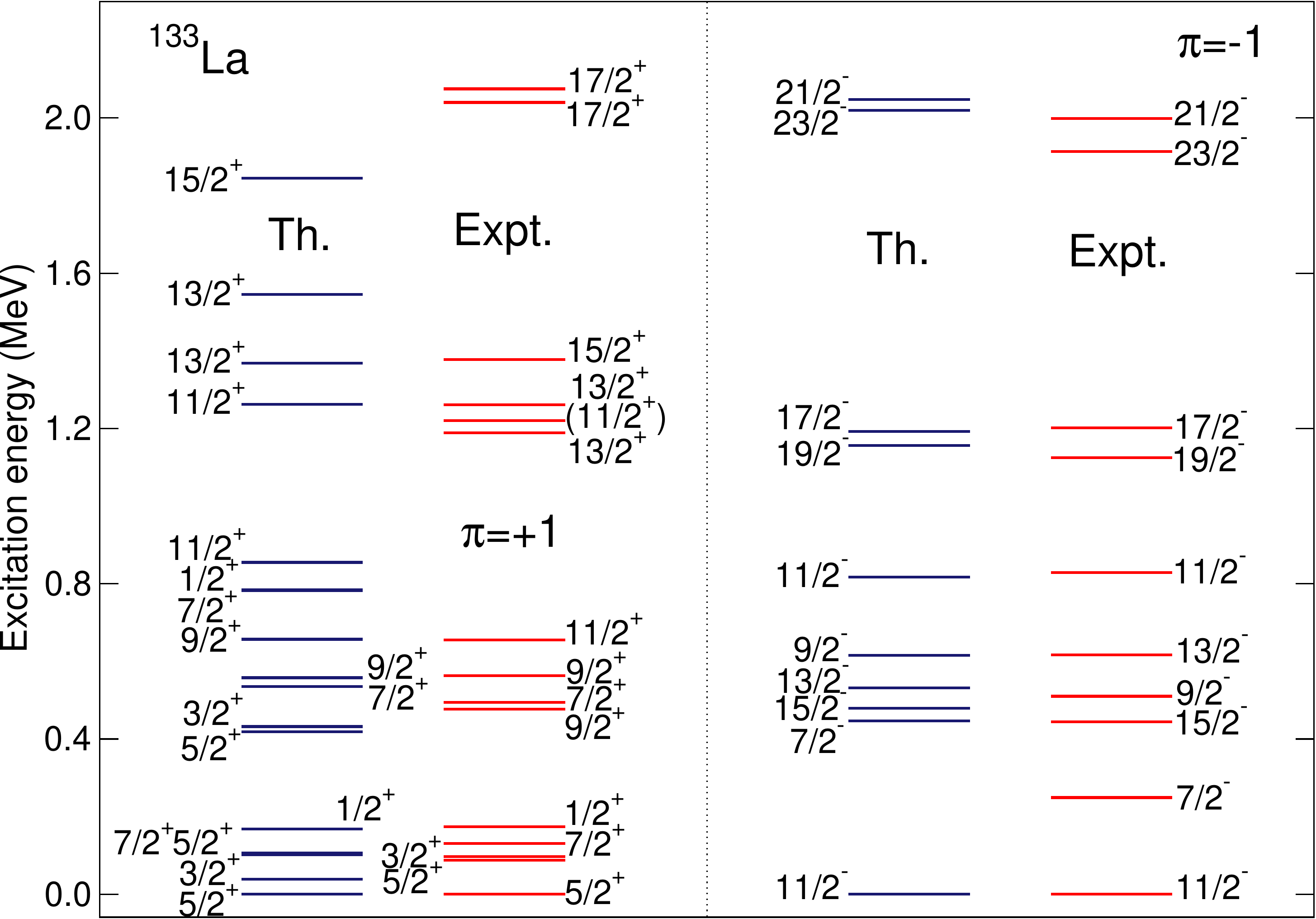}
\caption{(Color online) The  same as in Fig.~\ref{fig:135ba}, but for $^{133}$La.}
\label{fig:133la}
\end{center}
\end{figure}

% ----------------------------------------------------------------------
%                                                      T a b l e   V I
% ----------------------------------------------------------------------

\begin{table}[htb!]
\caption{\label{tab:133la}
The same as in Table~\ref{tab:135ba}, but for $^{133}$La.
}
\begin{center}
\begin{tabular}{p{2.0cm}cccc}
\hline\hline
\multirow{2}{*}{} & \multicolumn{2}{c}{$B(E2)$ (W.u.)} &
 \multicolumn{2}{c}{$B(M1)$ (W.u.)} \\
\cline{2-3} 
\cline{4-5}
          & Th.         & Expt.    & Th.         & Expt.     \\
\hline
${1/2}^+_{1}\rightarrow {3/2}^+_{1}$ & 3.4 & 6(3) & 0.68 & 0.017(6) \\
${1/2}^+_{1}\rightarrow {5/2}^+_{1}$ & 30 & 0.8(3) &  &  \\
${3/2}^+_{1}\rightarrow {5/2}^+_{1}$ & 32 & $>$35 & 0.083 & $>$0.026 \\
${5/2}^+_{2}\rightarrow {5/2}^+_{1}$ & 12 & 2.1(10) & 0.14 & 0.0097(8) \\
${7/2}^+_{1}\rightarrow {5/2}^+_{1}$ & 24 & 11(4) & 0.00013 & 0.0052(9) \\
${7/2}^+_{1}\rightarrow {5/2}^+_{2}$ & 21 & 6.1(20) & 5.9$\times 10^{-5}$ & 0.00068(16) \\
\hline\hline
\end{tabular}
\end{center}
\end{table}

% ----------------------------------------------------------------------
%                                            F i g u r e   1 9
% ----------------------------------------------------------------------

\begin{figure}[htb!]
\begin{center}
\includegraphics[width=\linewidth]{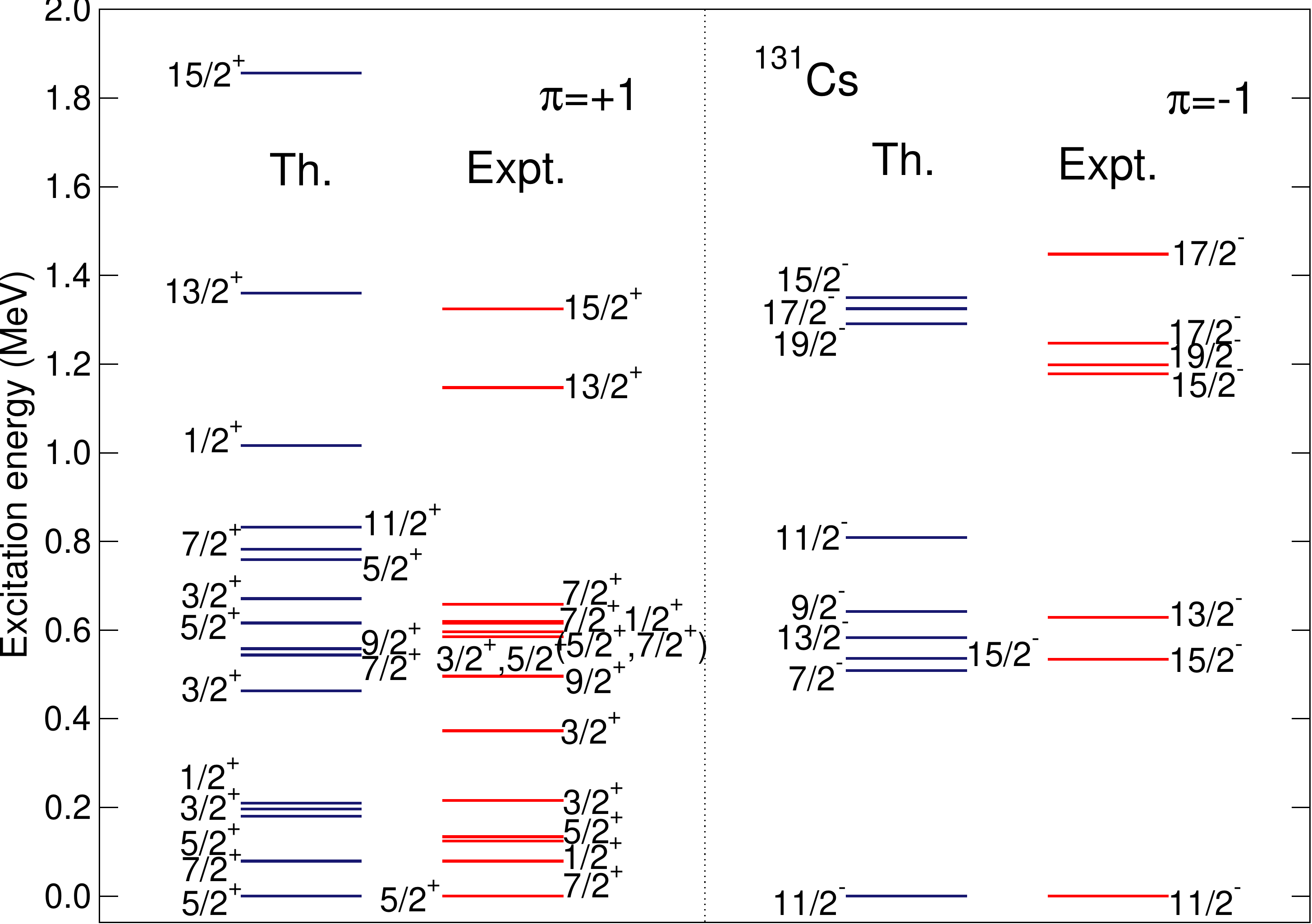}
\caption{(Color online) The  same as in Fig.~\ref{fig:135ba}, but for $^{131}$Cs.}
\label{fig:131cs}
\end{center}
\end{figure}

% ----------------------------------------------------------------------
%                                                      T a b l e   V I I
% ----------------------------------------------------------------------

\begin{table}[htb]
\caption{\label{tab:131cs}%
The same as in Table~\ref{tab:135ba}, but for $^{131}$Cs.}
\begin{center}
\begin{tabular}{p{2.0cm}cccc}
\hline\hline
\multirow{2}{*}{} & \multicolumn{2}{c}{$B(E2)$ (W.u.)} &
 \multicolumn{2}{c}{$B(M1)$ (W.u.)} \\
\cline{2-3} 
\cline{4-5}
          & Th.         & Expt.    & Th.         & Expt.     \\
\hline
${1/2}^+_{1}\rightarrow {5/2}^+_{1}$ & 55 & 69.5(14) &  &  \\
${1/2}^+_{2}\rightarrow {1/2}^+_{1}$ &  &  & 0.076 & 0.0010613(4) \\
${1/2}^+_{2}\rightarrow {3/2}^+_{1}$ & 6.3 & 0.09(4) & 0.067 & $3.4\times 10^{-5}(10)$ \\
${1/2}^+_{2}\rightarrow {3/2}^+_{2}$ & 23 & $>$0.62 & 0.0045 & $>5.8\times 10^{-5}$ \\
${1/2}^+_{2}\rightarrow {5/2}^+_{1}$ & 0.29 & 0.028248(4) &  &  \\
${1/2}^+_{2}\rightarrow {5/2}^+_{2}$ & 7.2 & 0.13835(5) &  &  \\
${3/2}^+_{1}\rightarrow {1/2}^+_{1}$ & 4.9 & 9(5) & 0.35 & 0.00339(10) \\
${3/2}^+_{1}\rightarrow {5/2}^+_{1}$ & 28 & 0.6(6) & 0.13 & 0.00922(5) \\
${3/2}^+_{1}\rightarrow {5/2}^+_{2}$ & 1.4 & $>$3.9 & 0.023 & $>4.1\times 10^{-5}$ \\
${3/2}^+_{1}\rightarrow {7/2}^+_{1}$ & 2.8 & 2.36(3) &  &  \\
${3/2}^+_{2}\rightarrow {1/2}^+_{1}$ & 0.66 & 2.4(4) & 0.031 & 0.00057(4) \\
${3/2}^+_{2}\rightarrow {3/2}^+_{1}$ & 2.6 & $>$2.1 & 0.0057 & $>7.8\times 10^{-5}$ \\
${3/2}^+_{2}\rightarrow {5/2}^+_{1}$ & 1.5 & 2.4(9) & 0.0033 & 0.00064(20) \\
${3/2}^+_{2}\rightarrow {5/2}^+_{2}$ & 4.1 & 0.5(4) & 0.0089 & 0.00071(4) \\
${3/2}^+_{2}\rightarrow {7/2}^+_{1}$ & 33 & 0.2122(3) &  &  \\
${5/2}^+_{2}\rightarrow {5/2}^+_{1}$ & 0.49 & 3.5(3) & 0.017 & 0.000369(17) \\
${5/2}^+_{2}\rightarrow {7/2}^+_{1}$ & 44 & $>$62 & 0.0026 & $<2.1\times 10^{-5}$ \\
${7/2}^+_{1}\rightarrow {5/2}^+_{1}$ & 0.0078 & 0.64(24) & 0.0013 & 0.00170(5) \\
\hline\hline
\end{tabular}
\end{center}
\end{table}

% ----------------------------------------------------------------------
%                                            F i g u r e   2 0
% ----------------------------------------------------------------------

\begin{figure}[htb!]
\begin{center}
\includegraphics[width=\linewidth]{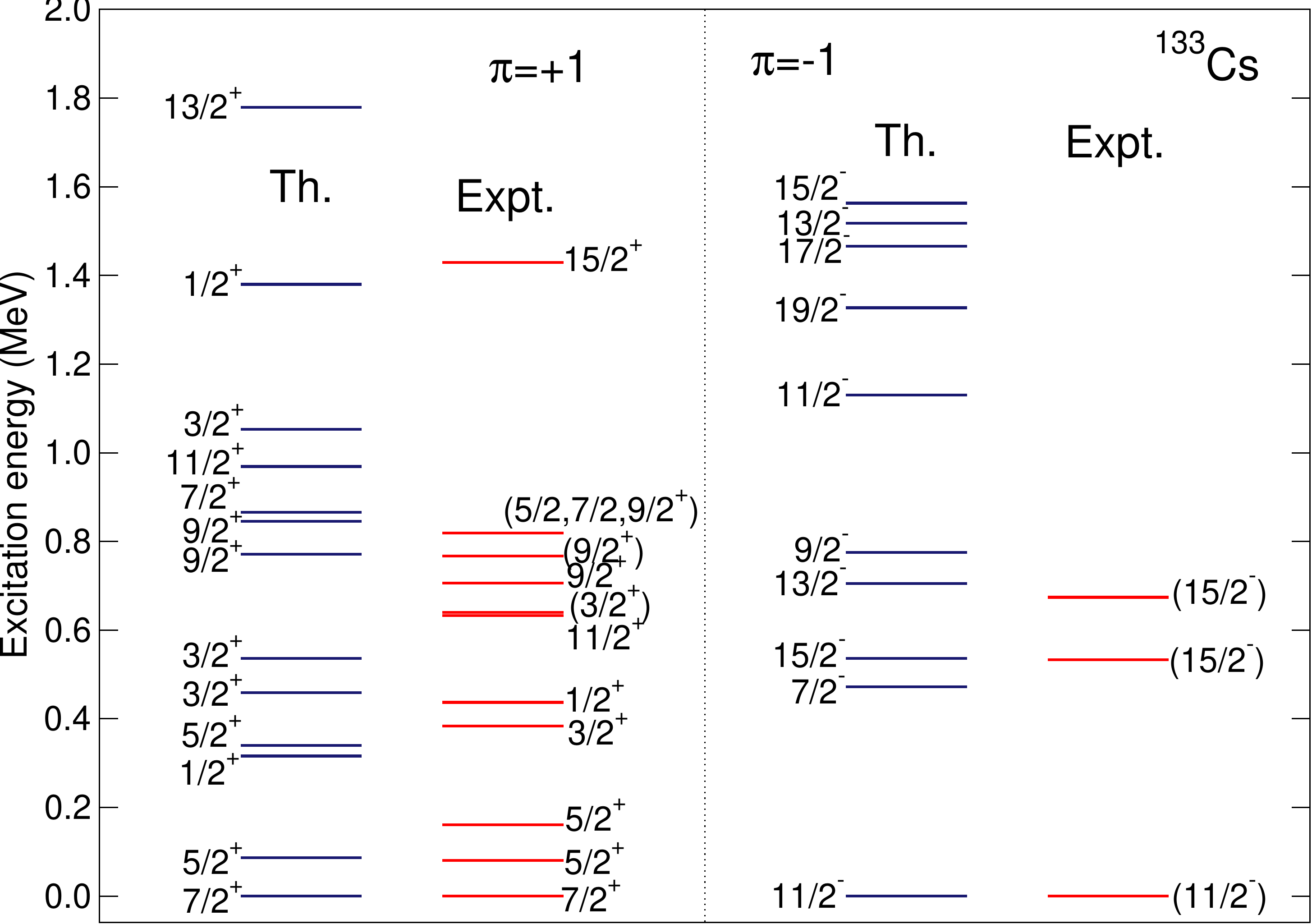}
\caption{(Color online) The same as in Fig.~\ref{fig:135ba}, but for $^{133}$Cs.}
\label{fig:133cs}
\end{center}
\end{figure}

% ----------------------------------------------------------------------
%                                                      T a b l e   V I I I
% ----------------------------------------------------------------------

\begin{table}[htb]
\caption{\label{tab:133cs}%
The same as in Table~\ref{tab:135ba}, but for $^{133}$Cs.}
\begin{center}
\begin{tabular}{p{2.2cm}cccc}
\hline\hline
\multirow{2}{*}{} & \multicolumn{2}{c}{$B(E2)$ (W.u.)} &
 \multicolumn{2}{c}{$B(M1)$ (W.u.)} \\
\cline{2-3} 
\cline{4-5}
          & Th.         & Expt.    & Th.         & Expt.     \\
\hline
${5/2}^+_{1}\rightarrow {7/2}^+_{1}$ & 0.49 & 5.8(4) & 0.00086 & 0.002381(22) \\
${5/2}^+_{2}\rightarrow {7/2}^+_{1}$ & 23 & 28.6(18) & 0.0022 & 0.081(4) \\
${5/2}^+_{2}\rightarrow {5/2}^+_{1}$ & 0.60 & 1.3$\times{}10^{2}(3)$ & 0.0024 & 0.000126(8) \\
${3/2}^+_{1}\rightarrow {5/2}^+_{2}$ & 0.024 & 0.12(4) & 0.020 & 0.00070(18) \\
${3/2}^+_{1}\rightarrow {5/2}^+_{1}$ & 17 & 0.04$^{+7}_{-4}$ & 0.12 & 0.011(3) \\
${3/2}^+_{1}\rightarrow {7/2}^+_{1}$ & 1.1 & 12(3) & & \\
${1/2}^+_{1}\rightarrow {3/2}^+_{1}$ & 8.8 & $>$18 & 0.75 & $>$0.024 \\
${1/2}^+_{1}\rightarrow {5/2}^+_{2}$ & 0.40 & $>$4.8 & & \\
${1/2}^+_{1}\rightarrow {5/2}^+_{1}$ & 43 & $>$12 & & \\
${11/2}^+_{1}\rightarrow {7/2}^+_{1}$ & 25 & 26.1(20) & & \\
${3/2}^+_{2}\rightarrow {5/2}^+_{2}$ & 2.7 & 3.6(23) & 0.0060 & 0.13(4) \\
${3/2}^+_{2}\rightarrow {5/2}^+_{1}$ & 2.0 & 1.0$\times{}10^{2}(7)$ & 0.0046 & 0.03(3) \\
${3/2}^+_{2}\rightarrow {7/2}^+_{1}$ & 26 & 3.4(7) & & \\
${9/2}^+_{2}\rightarrow {5/2}^+_{2}$ & 1.8 & 7.0(17) & & \\
${7/2}^+_{1}\rightarrow {7/2}^+_{3}$ & 18 & 6.5(18) & 0.039 & 0.016(4) \\
\hline\hline
\end{tabular}
\end{center}
\end{table}

\subsection{Detailed level schemes of selected odd-$A$ nuclei\label{sec:detail}}

In this section, we consider a selected set of odd-mass nuclei 
with available experimental data. For those nuclei
we present more detailed level schemes and 
electromagnetic transition rates. In particular, we will consider  
$^{135}$Ba, $^{129,131}$Xe, $^{133}$La and $^{131,133}$Cs
taken, as 
illustrative examples. We will also compare with the results 
obtained in Ref.~\cite{nomura2017odd-1}. The 
positive- and negative-parity energy spectra provided 
by our calculations are depicted 
in Figs.~\ref{fig:135ba}--\ref{fig:133cs}. The corresponding 
$B(E2)$ and $B(M1)$ values can be found in Tables~\ref{tab:135ba}--\ref{tab:133cs}.

Let us first discuss the odd-$N$ nuclei $^{135}$Ba and $^{129,131}$Xe, where
the low-lying states are based mainly on the $3s_{1/2}$, $2d_{3/2}$
(for positive-parity) and $1h_{11/2}$ (for negative-parity)
single-particle configurations. As can be seen from Fig.~\ref{fig:135ba}, exception made of the 
${3/2}^+_3$ and
${5/2}^+_3$ non-yrast states, the positive- and negative-parity
energy spectra obtained for $^{135}$Ba compare well with the 
experimental ones. Similar results have been obtained in Ref.~\cite{nomura2017odd-1}.
Furthermore, the results obtained in this study for the $B(E2)$ and $B(M1)$
transition probabilities are in a slightly better agreement
with the experimental data than those in Ref.~\cite{nomura2017odd-1}.

For both $^{129}$Xe (Fig.~\ref{fig:129xe}) and $^{131}$Xe
(Fig.~\ref{fig:131xe}),  we have obtained a good overall agreement
with the experiment. However, the predicted positive-parity levels are more
stretched than the experimental ones. Similar results have been found 
for $^{129}$Xe in Ref.~\cite{nomura2017odd-1}. From Tables~\ref{tab:129xe} and \ref{tab:131xe}
 one realizes, that the predicted
$B(E2)$ and $B(M1)$ transition rates compare well with
the experimental ones. However,  major discrepancies 
are observed for $B(E2; {5/2}^+_4\rightarrow
{1/2}^+_1)$ in $^{129}$Xe  and  
$B(E2; {7/2}^-_1\rightarrow
{11/2}^-_1)$ in $^{131}$Xe. The disagreement 
can be traced back to the structure of the 
IBFM wave functions. The ${1/2}^+_1$ and ${5/2}^+_4$ states in $^{129}$Xe are 
dominated by the $3s_{1/2}$ and $2d_{3/2}$ configurations, respectively,  leading 
to rather weak E2 transitions. On the other hand, the 
negative-parity states are
accounted for only by the $1h_{11/2}$ configuration.
Therefore, the rather small experimental $B(E2; {7/2}^-_1\rightarrow
{11/2}^-_1)$ transition probability in $^{131}$Xe might suggest that some additional
contributions, from outside of the employed IBFM model 
space, should still be taken into account.
Indeed, even  purely phenomenological IBFM calculations  \cite{ABUMUSLEH2014}
do not  reproduce the $B(E2; {7/2}^-_1\rightarrow
{11/2}^-_1)$ transition rate in $^{131}$Xe. Note, that the 
$B(E2)$ and $B(M1)$ transition rates  for $^{129}$Xe
are similar to the ones obtained in Ref.~\cite{nomura2017odd-1}.

Let us now discuss the odd-$Z$ nuclei $^{133}$La and
$^{131,133}$Cs. Their low-energy structures 
are mainly described  by the $1g_{7/2}$ and $2d_{5/2}$ (for
positive parity) and $1h_{11/2}$ (for negative parity) orbitals weakly
coupled to the corresponding IBM states.  
The  positive- and
negative-parity spectra of $^{133}$La are depicted in Fig.~\ref{fig:133la}. 
The comparison with the experiment reveals that, as in Ref.~\cite{nomura2017odd-1}, our calculations 
overestimate the positive-parity levels above 1 MeV excitation energy. 
On the other hand, exception made of the 
$B(E2; {1/2}^+_1\rightarrow {5/2}^+_1)$ and $B(E2;
{7/2}^+_1\rightarrow {5/2}^+_2)$ values, the computed E2 and M1
transitions  listed in
Table~\ref{tab:133la} agree reasonably well with the
experiment. Similar results have been obtained in Ref.~\cite{nomura2017odd-1}.

Finally, a reasonable agreement 
between the computed and the experimental spectra is observed 
in the case of $^{131}$Cs (Fig.~\ref{fig:131cs}) and $^{133}$Cs
(Fig.~\ref{fig:133cs}). As with $^{133}$La, the positive-parity
states with excitation energy above 1 MeV
are overestimated. 
The calculated $B(E2)$ and $B(M1)$ transition rates, shown in
Table~\ref{tab:131cs} (for $^{131}$Cs) and \ref{tab:133cs} (for
$^{133}$Cs), compare reasonably well with the experimental ones. 
Note, that the predicted E2 and M1 transition rates agree well with 
the results of Ref.~\cite{nomura2017odd-1}.

% ----------------------------------------------------------------------

\section{Summary and concluding remarks\label{sec:summary}}

% ----------------------------------------------------------------------

In this work, we have studied the spectroscopic properties of the odd-mass
$\gamma$-soft nuclei  $^{129-137}$Ba, $^{127-135}$Xe,  $^{129-137}$La
and $^{127-135}$Cs in terms of the IBFM with
the Hamiltonian constructed using the microscopic input provided 
by the Gogny-D1M HFB approximation. The
$(\beta,\gamma)$-deformation energy surfaces for the even-even core nuclei, 
spherical single-particle energies and occupation numbers for the
corresponding odd-mass systems have been computed within the
SCMF method, and these quantities have
been used to build the IBFM Hamiltonian \cite{nomura2016odd}. 
The coupling constants of the boson-fermion interaction terms have been 
fitted to experimental spectra. The  IBFM Hamiltonian has been then 
used to compute spectroscopic properties of the considered odd-mass systems.

For the even-even isotopes $^{128-136}$Ba and
$^{126-134}$Xe, the $(\beta,\gamma)$-deformation energy surfaces  suggest the
empirically observed structural evolution from  $\gamma$-soft to
nearly spherical shapes as functions of the neutron number. 
Our  (mapped) IBM calculations reproduce the experimental low-energy excitation
spectra in both even-even Ba and Xe nuclei reasonably well. 
Furthermore, the IBFM calculations provide a reasonable agreement 
with the available experimental data 
for the
low-energy positive- and negative-parity states as well as the
electromagnetic transition rates in the case of the odd-mass $^{129-137}$Ba,
$^{127-135}$Xe,  $^{129-137}$La and $^{127-135}$Cs nuclei. 
We have also shown that our results
are at the same level of accuracy in reproducing experimental data as
those reported in Ref.~\cite{nomura2017odd-1} on
the same set of the odd-mass nuclei but using the relativistic mean-field approximation.
The  results obtained in this study for $\gamma$-soft odd-mass nuclei, as well as the previous
ones for the axially-deformed odd-mass Eu and Sm systems
\cite{nomura2017odd-2}, further corroborate the validity of the employed method \cite{nomura2016odd}
based on the Gogny-EDF.

Even though the application of the method in its current version is
limited to only those nuclei where experimental data are available, the
results obtained in this work are promising and open up the possibility for
exploring odd-mass nuclear systems in other  regions of the nuclear chart. 
Our next step would be to apply the present method to neutron-rich
odd-mass nuclei, and a potential target are those with $A\approx
190-200$, including Pt and Os nuclei. They are expected
to display a rich variety of structural phenomena 
and sufficient experimental data
are already available. Indeed, the corresponding even-even nuclei in this mass
region exhibit prolate-oblate shape/phase transitions as well as 
examples of $\gamma$-softness. It would be, therefore, of interest to
study how the presence of an odd particle affects this type of shape/phase
transition. Work along this line is in progress, and will be reported
elsewhere.

\acknowledgments
K.N. acknowledges support from the Japan 
Society for the Promotion of Science. This work has been supported in 
part by the QuantiXLie Centre of Excellence. The  work of LMR was 
supported by Spanish Ministry of Economy and Competitiveness (MINECO)
Grants No. FPA2015-65929-P and FIS2015-63770-P.

\bibliography{refs}

\end{document}